\documentclass[letterpaper,fleqn,usenatbib]{mnras}
\pdfoutput=1
\usepackage{newtxtext,newtxmath}
\usepackage[T1]{fontenc}
\usepackage{ae,aecompl}

\usepackage{graphicx}	% Including figure files
\usepackage{amsmath}	% Advanced maths commands
\usepackage{amssymb}	% Extra maths symbols

\usepackage[dvipsnames]{xcolor}

\title[The diversity of the CGM]{The diversity of the circumgalactic medium around $z=0$ Milky Way-mass galaxies from the Auriga simulations}

\author[M. H. Hani et al.]{
  \parbox[t]{\textwidth}{
  {Maan H. Hani,$^{1}$\thanks{E-mail: mhani@uvic.ca}\thanks{Vanier Scholar}}
  {Sara L. Ellison,$^{1}$ }
  {Martin Sparre,$^{2,3}$ }   
  {Robert J. J. Grand,$^{4}$ }
  {R\"{u}ediger Pakmor,$^{4}$ }
  {Facundo A. Gomez,$^{5,6}$ }
  {Volker Springel$^{4}$}
  }
\\\\
$^{1}$Department of Physics and Astronomy, University of Victoria, Victoria, British Columbia, V8P 1A1, Canada\\
$^{2}$Institut f\"ur Physik und Astronomie, Universit\"at Potsdam, Karl-Liebknecht-Str.\,24/25, 14476 Golm, Germany\\
$^{3}$Leibniz-Institut f\"ur Astrophysik Potsdam (AIP), An der Sternwarte 16, 14482 Potsdam, Germany\\
$^{4}$Max-Planck-Institut f\"ur Astrophysik, Karl-Schwarzschild-Str. 1, D-85741 Garching, Germany\\
$^{5}$Instituto de Investigaci\'on Multidisciplinar en Ciencia y Tecnolog\'ia, Universidad de La Serena, Ra\'ul Bitr\'an 1305, La Serena, Chile\\
$^{6}$Departamento de F\'isica y Astronom\'ia, Universidad de La Serena, Av. Juan Cisternas 1200 Norte, La Serena, Chile\\
}

\date{Accepted XXX. Received YYY; in original form ZZZ}

\pubyear{2019}

\hypersetup{draft}
\begin{document}
\label{firstpage}
\pagerange{\pageref{firstpage}--\pageref{lastpage}}
\maketitle

%%%%%%%%%%%%%%%%%%%%%%%%%%%%%%%%%%%%%%%%%%%%%%%%%%

%%%%%%%%%%%%%%%%% BODY OF PAPER %%%%%%%%%%%%%%%%%%

\begin{abstract}
Galaxies are surrounded by massive gas reservoirs (i.e. the circumgalactic medium; CGM) which play a key role in their evolution.  The properties of the CGM, which are dependent on a variety of internal and environmental factors, are often inferred from absorption line surveys which rely on a limited number of single lines-of-sight.  In this work we present an analysis of 28 galaxy haloes selected from the Auriga project, a cosmological magneto-hydrodynamical zoom-in simulation suite of isolated Milky Way-mass galaxies, to understand the impact of CGM diversity on observational studies. Although the Auriga haloes are selected to populate a narrow range in halo mass, our work demonstrates that the CGM of L$^\mathrm{\star}$ galaxies is extremely diverse: column densities of  commonly observed species span $\sim 3-4$ dex and their covering fractions range from $\sim 5$ to $90$ per cent. Despite this diversity, we identify the following correlations: 1) the covering fractions (CF) of hydrogen and metals of the Auriga haloes positively correlate with stellar mass, 2) the CF of \ion{H}{i}, \ion{C}{iv}, and \ion{Si}{ii} anticorrelate with active galactic nucleus luminosity due to ionization effects, and 3) the CF of  \ion{H}{i}, \ion{C}{iv}, and \ion{Si}{ii} positively correlate with galaxy disc fraction due to outflows populating the CGM with cool and dense gas. The Auriga sample demonstrates striking diversity within the CGM of L$^\star$ galaxies, which poses a challenge for observations reconstructing CGM characteristics from limited samples, and also indicates that long-term merger assembly history and recent star formation are not the dominant sculptors of the CGM.
\end{abstract}


\section{Introduction}
\label{sec:intro}
\noindent
% The CGM and the big picture of galaxy formation 
The presence of an extended gas reservoir around galaxies is a fundamental prediction of the current paradigm of structure formation \citep{1978MNRAS.183..341W}. This gas reservoir, commonly referred to as the circumgalactic medium (CGM), arises as a natural consequence of gas collapse and acts as a liaison between a galaxy's interstellar medium (ISM) and the large-scale structure. 
% Evidence for the existence of the CGM
The existence of extended galactic gas haloes is borne out by a plethora of observations. Early studies of the CGM relied on serendipitous absorption in quasar spectra which was ascribed to foreground galaxies and their respective extended haloes \citep[e.g. ][]{1986A&A...155L...8B, 1991A&A...243..344B, 1995ApJ...442..538L, 2000ApJ...534L...1T, 2001ApJ...556..158C}. Recent studies of the CGM have placed our understanding on a more solid statistical and physical footing by studying statistically meaningful galaxy samples and systematically relating the CGM properties to the host galaxy properties \citep[e.g. ][]{2010ApJ...708..868C, 2011ApJ...740...91P, 2013ApJ...777...59T, 2014MNRAS.445.2061L, 2015ApJ...804...79L}. Contemporary CGM studies have thus demonstrated the ubiquitous nature of the CGM \citep[e.g. ][]{2012ApJ...758L..41T, 2013ApJS..204...17W, 2014ApJ...796..136B, 2017ApJ...837..169P}. Nonetheless, the observed CGM properties show a strong dependence on some galaxy properties such as stellar mass \citep{2011Sci...334..948T, 2014ApJ...796..136B}, colour \citep{2016ApJ...833..259B}, black hole activity \citep{2018MNRAS.478.3890B}, star formation activity \citep{2011Sci...334..948T, 2013ApJ...768...18B, 2017ApJ...846..151H, 2017ApJ...837..169P}, and environment \citep{2017ApJ...844...23P, 2018ApJ...869..153N}. For a review of the CGM, see \citet{2012ARA&A..50..491P}, and \citet{2017ARA&A..55..389T}.

% Properties of the CGM
Observational surveys imply that the CGM consists of a massive, metal-enriched, and clumpy multiphase gas reservoir with a declining density profile extending well beyond the virial radius of the host galaxy \citep[e.g. ][]{2013ApJS..204...17W, 2014ApJ...792....8W, 2014ApJ...788..119L, 2014MNRAS.445.2061L, 2015ApJ...804...79L, 2017ApJ...837..169P}.
% - clumpy multi-phase
The current picture of the CGM invokes a clumpy multiphase medium where cool/warm ($T\sim10^{4 - 5}$K) and dense gas clumps are embedded within a hot ($T \sim 10^6$K) and diffuse medium \citep[e.g. ][]{2009ApJ...698.1485H, 2017MNRAS.470..114A, 2017ApJ...848..122B}. The clumpy multiphase nature of the CGM is evident through the distinct observable ionization states: The cool dense clumps provide a suitable environment where the low ionization species can survive (e.g. \ion{H}{i}, \ion{Mg}{ii}, \ion{Si}{ii}, \ion{Si}{iii}, \ion{Si}{iv}, \ion{C}{iv}) while the hot diffuse medium hosts the highly ionized species (e.g. \ion{O}{vi}, \ion{O}{vii}). Additional evidence for a clumpy multiphase CGM arises from the analysis of quasar absorption line profiles \citep[e.g. ][]{2016ApJ...830...87S, 2016ApJ...833...54W} where gas clumps in the CGM contribute to the total absorption along a given line of sight whilst leaving a strong kinematic imprint on the absorption line profiles.

% - gas mass
Observational surveys targeting the CGM of L$^\star$ galaxies infer a CGM gas mass comparable to the mass of the ISM \citep[e.g. ][]{2010ApJ...714.1521C, 2011Sci...334..948T, 2014ApJ...792....8W, 2017ApJ...837..169P}, and report a positive correlation between CGM gas properties and properties of the ISM \citep{2015ApJ...813...46B}, suggesting a common evolution for both the ISM and the CGM. 
% - metal mass et metallicities
Additionally, studies focusing on the metal content of the CGM report a significant metal content \citep[e.g. ][]{2013ApJS..204...17W, 2014ApJ...786...54P}. For example, \citet{2014ApJ...786...54P} showed that, for L$^\star$ galaxies, a significant fraction of the metal budget resides in the CGM. The CGM's sizeable metal content is evident when examining CGM metallicities which may extend to supersolar metallicities \citep{2017ApJ...837..169P}: $\sim 25\%$ of the COS-Haloes sight lines exhibit metallicities which exceed the solar value. 

% Miky-Way CGM studies
The Milky Way provides a complementary approach to the large observational surveys that target the CGM of galaxy populations in a statistical fashion. Surveys of the Milky Way's CGM are often equipped with multiwavelength observations and a large number of lines-of-sight \citep[e.g. ][]{2012ARA&A..50..491P, 2015ApJ...799L...7F, 2017A&A...607A..48R}. Similar to other L$^\star$ galaxies, the Milky Way hosts a massive, extended, multiphase, and structurally/kinematically complex CGM \citep[e.g. ][]{ 2012ApJ...756L...8G, 2012ARA&A..50..491P, 2017ApJ...834..155M, 2017A&A...607A..48R, 2019arXiv190411014W, 2019ApJ...871...35Z}. Observations of the Galactic CGM infer a large covering fraction of gas thus indicating an extended and massive gas reservoir \citep{2012ApJ...756L...8G, 2017A&A...607A..48R, 2019ApJ...871...35Z}. The Galactic CGM also shows kinematically distinct features in \ion{H}{i} \citep[e.g. ][]{2017ApJ...834..155M} and low ionization species \citep[e.g. ][]{2017A&A...607A..48R, 2019arXiv190411014W} along different lines-of-sight which is indicative of a clumpy, multiphase medium. Whereas, the high-ionization (i.e. \ion{O}{vi}) gas is more uniform on larger scales \citep{2019arXiv190411014W}.

% Numerical simulations - a complementary approach
Numerical simulations of galaxy formation and evolution provide a complementary view to the results from observations. A particular strength of numerical simulations is that they provide an environment to study the evolution of the CGM and the feedback processes shaping its structure, ionization, and metallicity \citep[e.g. ][]{2013MNRAS.430.1548H, 2016MNRAS.462..307B, 2016ApJ...824...57C, 2016MNRAS.460.2157O, 2018MNRAS.477..450N}. Simulated CGM properties show many remarkable similarities to those observed.
% Numerical simulations re-poroduce obs
For example, recent cosmological hydrodynamical simulations reproduce the observed extended galactic gas haloes with a sizeable, metal-enriched, and multiphase gas reservoir \citep[e.g. ][]{2012MNRAS.425.1270S, 2013MNRAS.432...89F, 2013ApJ...765...89S, 2016MNRAS.460.2881N, 2017MNRAS.465.2966S, 2018A&A...609A..66M}. Additionally, numerical simulations demonstrate the redshift evolution of the CGM and its structure \citep[e.g. ][]{2012ApJ...760...50S,  2012MNRAS.421.2809V, 2014MNRAS.444.1260F, 2015MNRAS.454.2691M, 2016ApJ...824...57C,  2016ApJ...827..148C}. The observed bimodality in the \ion{O}{vi} column densities and covering fractions between star forming and quenched galaxies has also been addressed by different numerical simulations \citep[e.g. ][]{2015MNRAS.448..895S, 2016MNRAS.460.2157O, 2018MNRAS.477..450N}. 

% Strength of simulations
Various numerical works have emphasized the importance of galactic outflows for populating the CGM with metals out to large radii and driving some of the commonly observed CGM properties \citep[e.g. ][]{2006MNRAS.373.1265O, 2013MNRAS.430.1548H, 2015MNRAS.448..895S, 2016MNRAS.462..307B, 2018MNRAS.475.1160H}. For example, numerical simulations have demonstrated the vital role of outflows driven by supernova feedback at populating the CGM with metals: Metal-rich winds launched by supernovae (SN) deposit metals in the CGM, thus giving rise to the high metallicities reported by observations \citep[e.g. ][]{2006MNRAS.373.1265O, 2013MNRAS.430.1548H, 2014MNRAS.444.1260F,  2016MNRAS.462..307B}. In addition to being strong drivers of outflows \citep[e.g. ][]{2015MNRAS.448..895S, 2018MNRAS.477..450N}, feedback from active galactic nuclei (AGN) has a profound effect on the ionization structure of the CGM through an aggressive radiation field which keeps the CGM out of ionization equilibrium \citep[e.g. ][]{2013MNRAS.434.1063O, 2017MNRAS.471.1026S, 2018MNRAS.481..835O, 2018MNRAS.474.4740O}. Additionally, the CGM is shaped by small-scale physical processes (e.g. turbulent and conductive mixing) which have significant effects on the CGM's large-scale structure; simulations modelling the CGM or clouds of gas at high resolution particularly underline the effects of resolution on the CGM and its neutral gas content in cosmological \citep[e.g. ][]{ 2018arXiv181105060C, 2019ApJ...873..129P,  2019MNRAS.483.4040S, 2019MNRAS.482L..85V} as well as idealized settings \citep[e.g. ][]{2015ApJ...805..158S, 2017MNRAS.470..114A, 2018arXiv181012925F, 2018MNRAS.473.5407M, 2018ApJ...862...56S, 2019MNRAS.482.5401S}. Furthermore, simulated galaxy interactions stress the long lasting effects an interaction can have on the CGM metal content and temperature/ionization \citep[e.g. ][]{2006ApJ...643..692C, 2018MNRAS.475.1160H}.

The framework in which the CGM is driven by feedback processes on a galactic scale (i.e. AGN and SN feedback) is reflected by the ubiquitous observations of outflows and their effects on the CGM kinematics, densities, and ionization. Observational studies show that vigorous outflows arising from enhanced star formation \citep[e.g. ][]{2005ApJ...621..227M, 2005ApJS..160..115R, 2009ApJ...697.2030S} or AGN activity \citep[e.g. ][]{2005ApJ...632..751R, 2016ApJ...832..142Z, 2017ApJ...839..120W} can populate the CGM with metals and alter its density structure, thus giving rise to the observed multiphase absorbers \citep[e.g. ][]{2013ApJ...768...18B, 2015MNRAS.447.1834B, 2016ApJ...822....9H, 2017ApJ...848..122B, 2017ApJ...846..151H}. Consequently, starburst galaxies and AGN hosts show different CGM properties when compared to their non-active counterparts \citep{2017ApJ...846..151H, 2018MNRAS.478.3890B}. The effects of outflows have also been demonstrated by studies focusing on the CGM where low ionization species exhibit an azimuthal dependence in disc-dominated galaxies \citep[e.g. ][]{2011ApJ...743...10B, 2017ApJ...834..148N}. The Milky Way's CGM provides a local illustration of outflows and their influence on the CGM. For example, the Fermi Bubbles are the $\gamma$-ray \citep{2010ApJ...724.1044S} signature of a biconical outflow originating at the centre of the Galaxy where hot gas sweeps the cool gas into the flow \citep[i.e. ][]{2006ApJ...646..951K, 2008ApJ...679..460Z, 2015ApJ...799L...7F}. In addition to the effects of galactic feedback, a galaxy's environment drives changes in the CGM structure, for example the CGM of galaxies in group environments exhibits distinct kinematics \citep{2017ApJ...844...23P, 2018ApJ...869..153N} and covering fractions \citep[e.g. ][]{2015MNRAS.449.3263J, 2016ApJ...832..124B}. 

Taken together, observations and simulations show that whilst the \textit{presence} of a CGM is very common, in its details, these galactic gas reservoirs are both complex and sensitive to a number of structural (e.g. total stellar or halo mass, and morphology), internal (star formation and nuclear activity), and environmental (clustering and merging) properties.  However, many of these properties are interrelated.  For example, high-mass galaxies tend to have more bulge-dominated morphologies, more massive central black supermassive black holes (and more frequent AGN activity), and higher star formation rates. Empirical studies of the CGM are usually reliant on single lines of sight that cannot capture the diversity of the CGM within a given galaxy. Therefore, numerical simulations can provide a valuable tool to quantify the impact of this diversity and the effects of limited observational samples.

In the work presented here, we study the properties of the CGM in a set of simulations designed to yield $z=0$ Milky Way-mass galaxies (hereafter referred to as L$^\star$ galaxies). These simulated galaxies therefore all have a consistent endproduct --galaxies similar to the Milky Way-- but they show significant variation in the rest of their properties, such as morphology \citep[i.e. ][]{2016MNRAS.459..199G, 2017MNRAS.466.3859M}, star formation rate and AGN luminosity. The goal of this work is therefore twofold: first to quantify the diversity of line-of-sight column densities through a given L$^\star$-like halo, and second, to study how CGM properties correlate with the host properties and therefore identify the mechanisms responsible for shaping the CGM. This paper is structured as follows: in section \ref{sec:methods} we describe the numerical model, the halo selection method, and the ionization analysis. We then present the results of the simulations in section \ref{sec:results} highlighting the major correlations between the host properties and the CGM ionization and metal content. In section \ref{sec:discussion} we discuss the implications of the diversity in the CGM on observational survey results. Lastly, section \ref{sec:conclusions} summarizes the conclusions of this work.
\section{Methods}
\label{sec:methods}
\noindent
The work presented here aims to understand the processes responsible for shaping the CGM of $z = 0$, Milky Way-mass galaxies. We use cosmological zoom-in simulations that provide a self-consistent cosmological framework to study the CGM while allowing for superior spatial and mass resolutions when compared to large-box cosmological simulations. The simulations used in this study are part of the Auriga simulation suite \citep{2017MNRAS.467..179G}. The simulated galaxies reproduce several observable properties -- such as disc galaxies with appropriate stellar masses, stellar disc sizes, rotation curves, star formation rates, and metallicities \citep[see ][]{2017MNRAS.467..179G} -- which makes the Auriga suite ideal to study the CGM of Milky Way-mass galaxies in the local universe.

\begin{figure*}
\centering
\includegraphics[width=\textwidth]{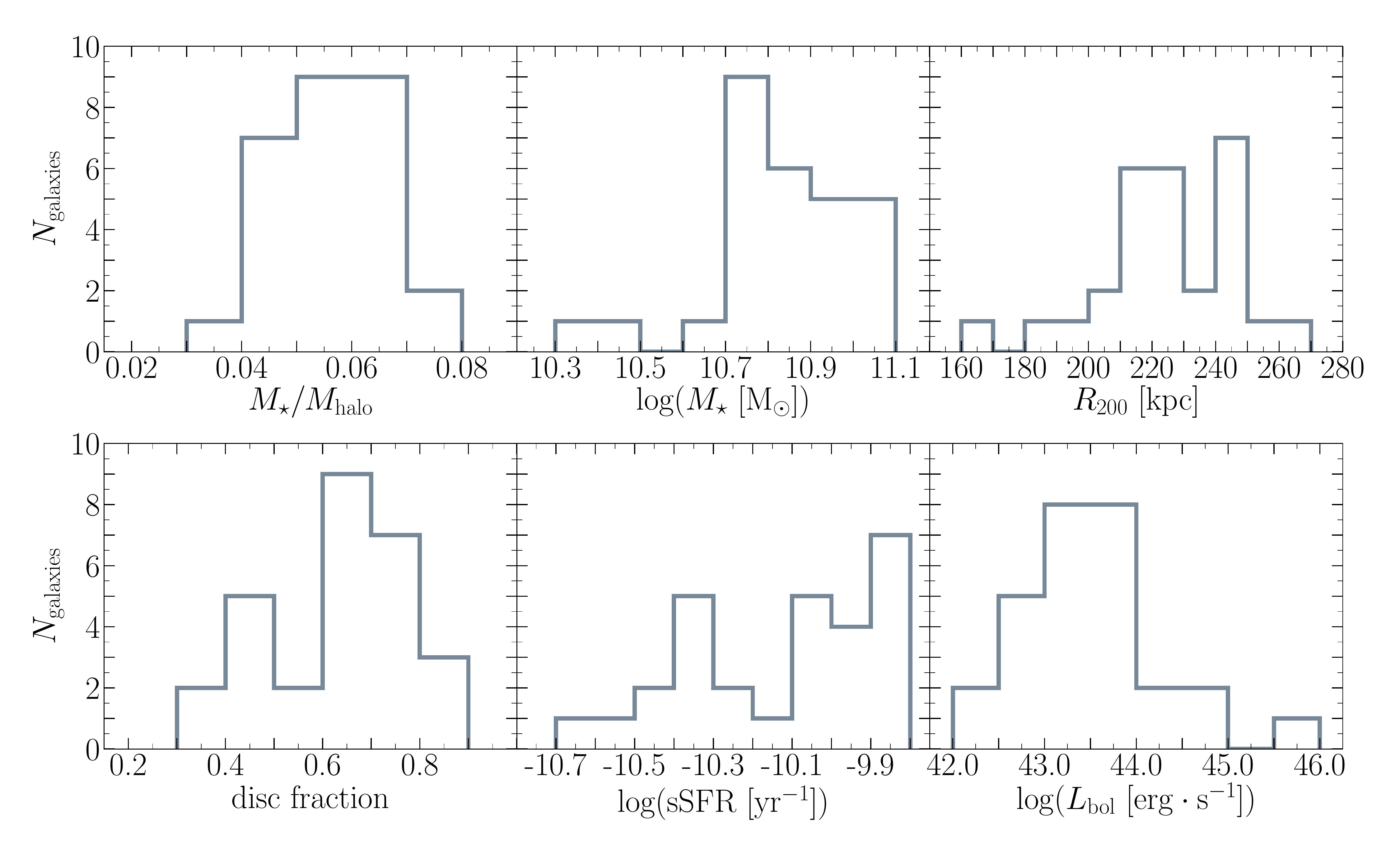}
\caption{A summary of the galaxy properties for the 28 Auriga haloes studied in this work: stellar mass to halo mass ratio ($M_\star / M_\mathrm{halo}$), stellar mass ($M_\star$), virial radius ($R_{200}$), disc fraction, specific star formation rate ($\mathrm{sSFR}$), and bolometric AGN luminosity ($L_\mathrm{bol}$).}
\label{fig:sample_properties}
\end{figure*}

% --------------------------------------------------------------------------------
% --------------------------------------------------------------------------------
% Auriga suite, extension, physics and specific halo selection.
% --------------------------------------------------------------------------------
% --------------------------------------------------------------------------------
\subsection{The Auriga simulation suite and halo selection}
\label{sec:methods-auriga}
\noindent
% What is auriga
The original Auriga simulation suite consists of 40\footnote{Ten haloes have been added to the original $30$ since the suite was introduced in \citet{2017MNRAS.467..179G}} magneto-hydrodynamical, cosmological zoom-in simulations of L$^\star$ haloes (Milky Way-like in halo mass) in isolated environments. The galaxies are simulated using the quasi-Lagrangian moving-mesh code \textsc{arepo} \citep{2010MNRAS.401..791S, 2016MNRAS.455.1134P} with dark matter resolution elements of mass $M_\mathrm{dm}\sim 3\times 10^{5}\, \mathrm{M_\odot}$, and a gas mass resolution of $M_\mathrm{gas} \sim 5\times 10^{4}\, \mathrm{M_\odot}$ (i.e. resolution level 4 as defined by \citealt{2017MNRAS.467..179G}). The Auriga project adopts a $\Lambda$CDM cosmology where $\Omega_\mathrm{m} = 0.307$, $\Omega_\mathrm{b} = 0.048$, $\Omega_\Lambda = 0.693$, and $H_0 = 67.77 \, \mathrm{km \, s^{-1}\, Mpc^{-1}}$.

% The physics model 
The Auriga simulation suite employs a comprehensive galaxy formation model \citep{2017MNRAS.467..179G} which is based on the physical model in the Illustris project \citep{2013MNRAS.436.3031V} and the modifications of \citet{2014MNRAS.437.1750M}. The model is described in detail in \citet{2017MNRAS.467..179G}. Here, we present a summary of the galaxy formation model which includes: 
\begin{enumerate}
\item{an effective equation of state describing a multiphase ISM in pressure equilibrium \citep{2003MNRAS.339..289S},}
\item{a star formation model and associated stellar feedback in the form of a phenomenological wind model, as well as metal enrichment from asymptotic giant branch (AGB) stars, SN type Ia, and SN II \citep{2013MNRAS.436.3031V, 2014MNRAS.437.1750M, 2017MNRAS.467..179G},}
\item{a model accounting for black hole (BH) seeding, growth, and associated feedback from AGN in radio and quasar modes \citep{2005MNRAS.361..776S,  2013MNRAS.436.3031V, 2014MNRAS.437.1750M, 2017MNRAS.467..179G} using the scaling of \citet{2000MNRAS.311..346N},}
\item{gas cooling via primordial channels alongside metal-line cooling \citep{2013MNRAS.436.3031V},}
\item{gas heating from a spatially uniform, redshift-dependent ultraviolet background \citep[UVB; ][]{2009ApJ...703.1416F} and radiation from nearby AGN \citep{2013MNRAS.436.3031V, 2017MNRAS.467..179G}, and}
\item{a prescription for magnetic fields \citep{2013MNRAS.432..176P, 2014ApJ...783L..20P}.}
\end{enumerate}
The Auriga model deviates from the Illustris model \citep{2013MNRAS.436.3031V} in two major ways: (1) the Auriga model implements a gentler radio-mode AGN feedback \citep{2014MNRAS.437.1750M} where several hot buoyant bubbles are inflated stochastically, and isotropically in the halo (following an inverse square distance profile) each with a fraction of the feedback energy prescribed in \citet{2013MNRAS.436.3031V} thus yielding much gentler halo heating, and (2) \textit{isotropic} stellar winds \citep{2017MNRAS.467..179G} are also endowed by a thermal energy budget thus reducing the mass loading \citep{2014MNRAS.437.1750M}.

% Extension galaxies and selection
In the work presented here, we follow stringent selection criteria to ensure that the sample only includes galaxies which have had quiet recent merging histories (no mergers more major than $1:3$ within the past $2$ Gyr) and is therefore free of contaminants (i.e. ongoing interactions and post-merger galaxies) which may skew the results. Knowing that recent galaxy interactions can have a noticeable impact on the CGM \citep{2018MNRAS.475.1160H}, we exclude\footnote{We exclude Auriga haloes Au8, Au11, Au21, Au25, Au30, Au$^\mathrm{x}$2, Au$^\mathrm{x}$3, Au$^\mathrm{x}$4, Au$^\mathrm{x}$5, Au$^\mathrm{x}$6, Au$^\mathrm{x}$9, and Au$^\mathrm{x}$10} some of the Auriga haloes from our analysis based on:
\begin{enumerate}
\item{the existence of a neighbouring galaxy more massive than $1/3 \times M_\star$ of the central halo and at a distance $R<0.4\times R_\mathrm{vir}$ from the central halo, or}
\item{the presence of a recent (within the past $2$ Gyr) major merger (more major than $1:3$) based on the merger trees, or}
\item{the existence of prominent signatures (visually inspected) of a recent or ongoing interaction in the stellar or gas profiles (i.e. tidal tails, caustic structures, streams).}
\end{enumerate}
\noindent 
Our selection criteria yield a sample of 28 Auriga galaxies with quiet recent merger histories. The haloes studied in this work are all at $z=0$, and they span a limited range in halo mass, $ 0.5 \leq M_{200}\, [10^{12}\,\mathrm{M}_\odot] \leq 2.0$, and stellar mass, $ 2.1 \leq M_\star\, [10^{10} \,\mathrm{M}_\odot] \leq 11.7$. The selected haloes are actively star forming with star formation rates $1.9 \leq \mathrm{SFR}\,[\mathrm{M}_\odot \mathrm{yr}^{-1}] \leq 12.4$, and AGN bolometric luminosities $10^{42.1} \leq L_\mathrm{bol} \,[\mathrm{erg s^{-1}}] \leq 10^{45.5}$. Figure \ref{fig:sample_properties} provides a summary of the galactic properties of the sample studied in this work. 

The complete Auriga suite includes AGN with bolometric luminosities $\ge 41.7\,\mathrm{erg s^{-1}}$ consequently causing the subsample we selected to have bolometric luminosities $\ge 42.1\,\mathrm{erg s^{-1}}$. Additionally, the Auriga suite produces star forming galaxies at $z=0$ \citep[see ][]{2017MNRAS.467..179G}. We note that the star formation rates and AGN luminosities in our sample are higher than comparable observational studies focusing on Milky Way-mass galaxies \citep[e.g. ][]{2012ApJS..198....3W}. Other studies with comparably high SFRs or $L_\mathrm{bol}$ explicitly select their sample to control for the effects of star formation \citep[e.g., ][]{2011Sci...334..948T}, starbursts \citep[i.e. COS-BURST; ][]{2017ApJ...846..151H}, or AGN \citep[i.e. COS-AGN; ][]{2018MNRAS.478.3890B}. One should proceed with caution when quantitatively comparing the results reported here to observational studies of the CGM. The results presented in the following sections investigate the physical processes responsible for shaping the CGM within the Auriga model. Hence, the qualitative correlations are not affected by the high SFR and $L_\mathrm{bol}$ values.

% --------------------------------------------------------------------------------
% --------------------------------------------------------------------------------
% Gas ionisation calculations
% --------------------------------------------------------------------------------
% --------------------------------------------------------------------------------
\subsection{CGM ionization}
\label{sec:methods-ionization}
\noindent
% ionisation calculation
Unlike metals (i.e. C, N, O, Ne, Mg, Si, Fe) which are explicitly tracked in the simulations, ionization species are not explicitly tracked on-the-fly in the simulations. Therefore, post-processing of the simulation results is required to calculate the ionization balance of gas cells. In this work, we follow the methodology introduced in \citet{2018MNRAS.475.1160H} where a grid of single cell \textsc{cloudy} \citep{2013RMxAA..49..137F} simulations is interpolated to calculate the ionization balance in each gas cell depending on the cell's temperature, density, metallicity, the bolometric AGN intensity at the cell's location, and the redshift of the galaxy \citep[see table 2 in][]{2018MNRAS.475.1160H}. The ionization calculations account for contributions from two radiation fields: an AGN radiation field, and a spatially uniform time varying UVB \citep{2009ApJ...703.1416F}, and assume ionization equilibrium. The effects of self-shielding are also incorporated following \citet{2013MNRAS.430.2427R}.

% --------------------------------------------------------------------------------
% --------------------------------------------------------------------------------
% LOS et covering fractions
% --------------------------------------------------------------------------------
% --------------------------------------------------------------------------------
\subsection{Tracking CGM properties}
\label{sec:methods-cf}
\noindent
% covering fraction calculation
We analyse the CGM of each Auriga halo within three square fields of view (FOVs) along three random\footnote{
We align the viewing angles with the simulation box axes. The box coordinates do \textit{not} align with the haloes' angular momentum vectors. Therefore, viewing the galaxies along the box axes is synonymous to using random viewing angles.} viewing angles. The FOVs extend out to the virial radius of the central galaxy ($R_{200}$). Namely, each FOV has a size of $2R_\mathrm{vir} \times 2R_\mathrm{vir}$. The normalization of the FOVs to the galaxies' virial radii ensures a consistent comparison between galaxies regardless of stellar or halo mass. See Figure \ref{fig:sample_properties} for the distribution of virial radii: the median $R_\mathrm{vir} = 225 \, \mathrm{kpc}$ yields an FOV of $450\times 450\, \mathrm{kpc}^2$.

Each FOV is uniformly sampled by $1024^2$ lines-of-sight with a depth of $6R_\mathrm{vir}$. Sampling the CGM below the simulation's inherent spatial resolution (i.e. oversampling) allows us to compute the `true' covering fraction because the surface area covered by the gas converges when sampling at a higher resolution than that of the underlying gas field. See Section \ref{sec:discussion} for a discussion of the effects of undersampling the CGM.

We compute the column densities of four ionic species, i.e. \ion{H}{i}, \ion{Si}{ii}, \ion{C}{iv}, \ion{O}{vi}, and their parent elements (i.e. H, Si, C, O) along each line-of-sight. Subsequently, covering fractions of each ion/element above a limiting column density threshold are calculated for each Auriga halo using all the lines-of-sight from the three FOVs (i.e., $3 \times 1024^2$ LOSs per CGM). In this work we are primarily interested in the absorption properties of the CGM (within the virial radius), rather than the properties of the galactic disc's ISM. Therefore, we ignore lines-of-sight with impact parameters $\le 0.2 R_\mathrm{vir}$ in all covering fraction calculations.

Knowing that the focus of this study is to understand the effects of galaxy evolution on the $z=0$ CGM, we focus our analysis on the $z=0$ snapshots. The evolutionary histories are only used when selecting the sample for this study. 

\section{Results}
\label{sec:results}
\noindent
Our current understanding of galaxy evolution draws strong connections between evolutionary processes and their effect on observable galaxy properties \citep[for a review see ][]{2013ARA&A..51..511K}. For example, the evolution and impact of BHs has been tied to the bulge properties of the host galaxy where it is thought that BH feedback, and processes which feed the central BH drive the $M_\mathrm{BH}-\sigma_\mathrm{bulge}$ relationship \citep[e.g. ][]{2013ApJ...764..184M}. The effects of BHs extend beyond the host galaxy and are sometimes seen as inflated bubbles in the intergalactic medium \citep{2012ARA&A..50..455F}. Similarly, galaxy mergers are often thought to be strong drivers of changes in galaxy morphology \citep[e.g. ][]{1987ApJ...312....1H, 1996ApJ...471..115B, 2008MNRAS.391.1137L, 2014MNRAS.445.1157C, 2016MNRAS.461.2589P,  2017MNRAS.470.3946S, 2018MNRAS.480.1715P}.

This work aims at understanding the processes that shape the CGM of L$^\star$ galaxies. Therefore, we focus our analysis on understanding the physical processes driving correlations between the CGM and `observed' galactic properties of the simulated Auriga galaxies.

% --------------------------------------------------------------------------------
% --------------------------------------------------------------------------------
% CGM diversity
% --------------------------------------------------------------------------------
% --------------------------------------------------------------------------------
\subsection{The diverse CGM}
\label{sec:results-diversity}
\noindent
The Auriga haloes studied here are selected to have similar halo mass and recently quiet histories (see Section \ref{sec:methods-auriga}). However, they none the less have a range of SFRs, disc fractions, and AGN activity (see Figure \ref{fig:sample_properties}). In this section, we explore the interhalo and intrahalo diversity of the CGM in the Auriga sample.

\begin{figure}
\centering
\includegraphics[width=\columnwidth]{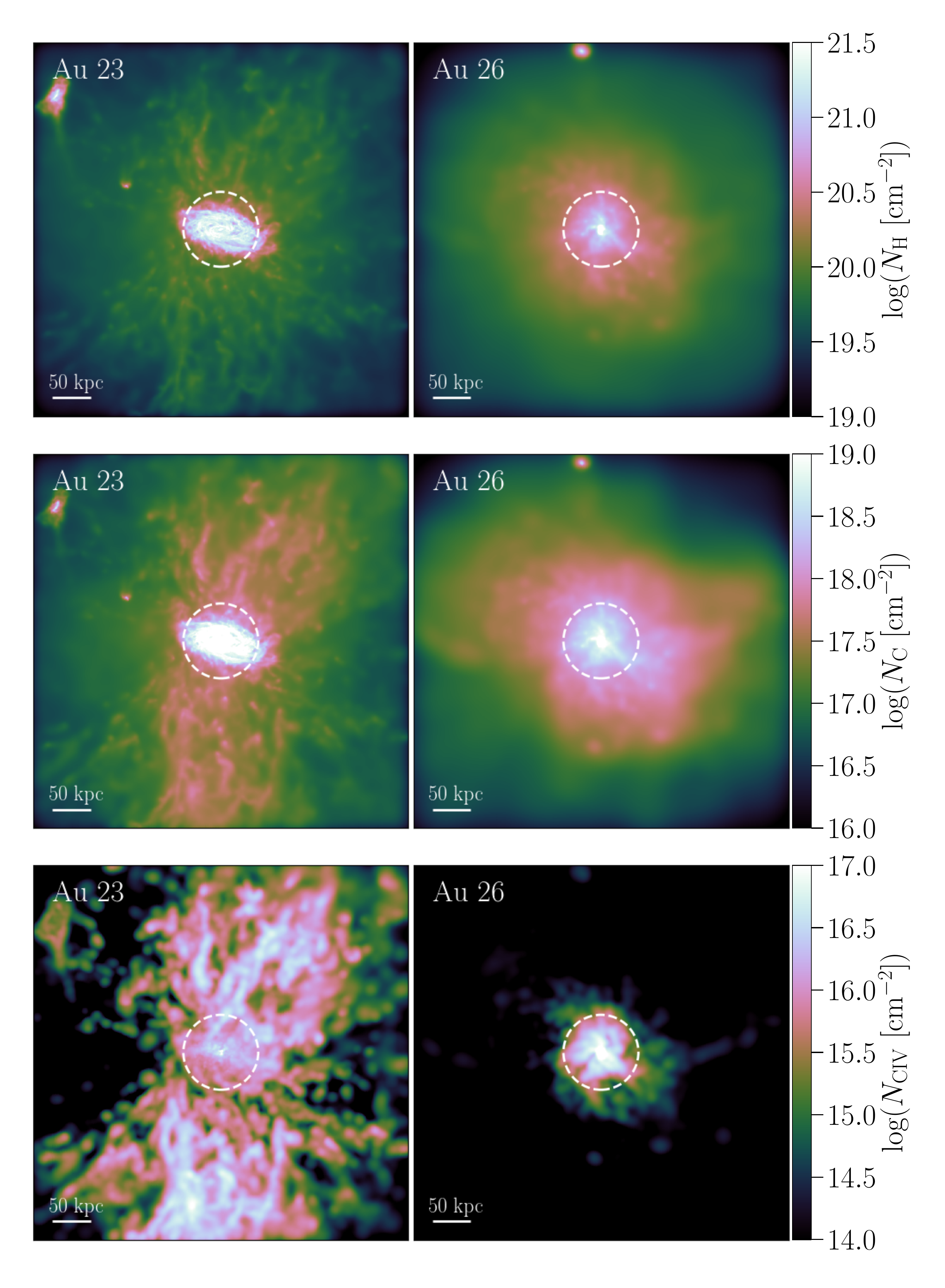}
\caption{Sample projections ($2R_{200} \times 2R_{200}$) of hydrogen (top), carbon (middle), and \ion{C}{iv} (bottom) for Au 23 (left-hand panel) and Au 26 (right-hand panel). The two haloes show significant differences in their covering fractions of \ion{C}{iv} as well as their CGM structure. While Au 26 shows a smooth CGM, the CGM of Au 23 is rich in filaments and clumps. Additionally, the signature of conical outflows is evident in Au 23.}
\label{fig:projection}
\end{figure}

Figure \ref{fig:projection} shows projections of two sample Auriga haloes, Au 23 (left-hand panels) and Au 26 (right-hand panels), in hydrogen (top panels), carbon (middle panels) and \ion{C}{iv} (bottom panels). The two haloes depicted in Figure \ref{fig:projection} show significant diversity in their CGM structure, column density distributions, and covering fractions of carbon and \ion{C}{iv}. While Au 23 exhibits strong CGM structures (i.e. filaments, clumps), the CGM of Au 26 is smoothly varying which is evident in the \ion{C}{iv} projection.

\begin{figure*}
\centering
\includegraphics[width=\textwidth]{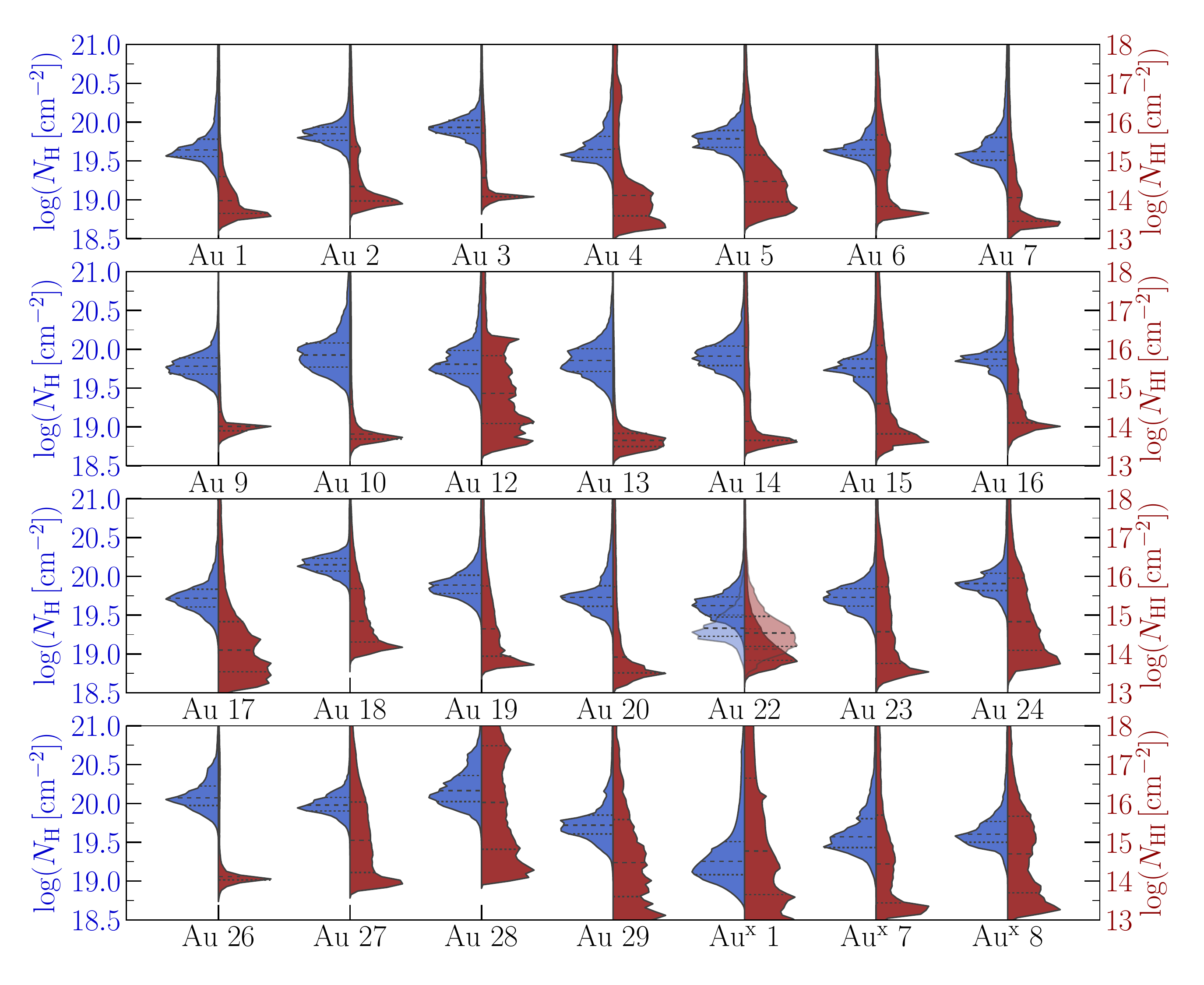}
\caption{A grouped split violin plot representing the column density distributions of hydrogen (blue, left) and \ion{H}{i} (red, right) along lines-of-sight (one viewing angle) through the 28 Auriga haloes. The dashed lines indicate the median column density of each distribution while the interquartile range is indicated by the dotted lines. The transparent violin plot represents the re-run of Au 22 with no AGN feedback (i.e. Au 22-NOAGN). The diversity in the CGM of each halo is evident by the column density distributions that extend over several decades in column density. Additionally, the diversity in CGM properties within the Auriga sample is evident by the vast differences between the haloes in the shapes and medians of the distributions.}
\label{fig:los-distro-H}
\end{figure*}

\begin{figure*}
\centering
\includegraphics[width=\textwidth]{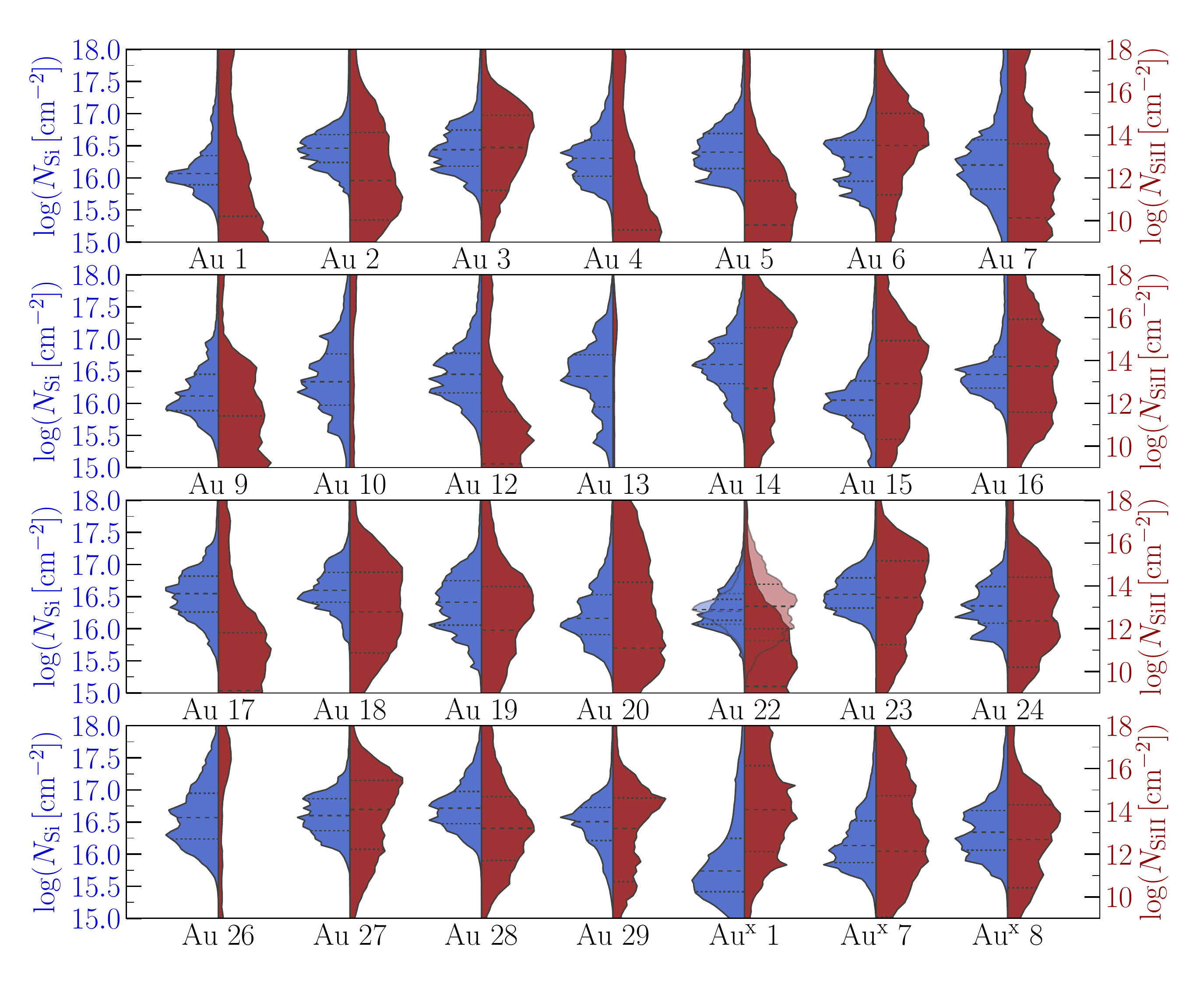}
\caption{A grouped split violin plot representing the column density distributions of silicon (blue, left) and \ion{Si}{ii} (red, right) along lines-of-sight (one viewing angle) through the 28 Auriga haloes.}
\label{fig:los-distro-Si}
\end{figure*}

\begin{figure*}
\centering
\includegraphics[width=\textwidth]{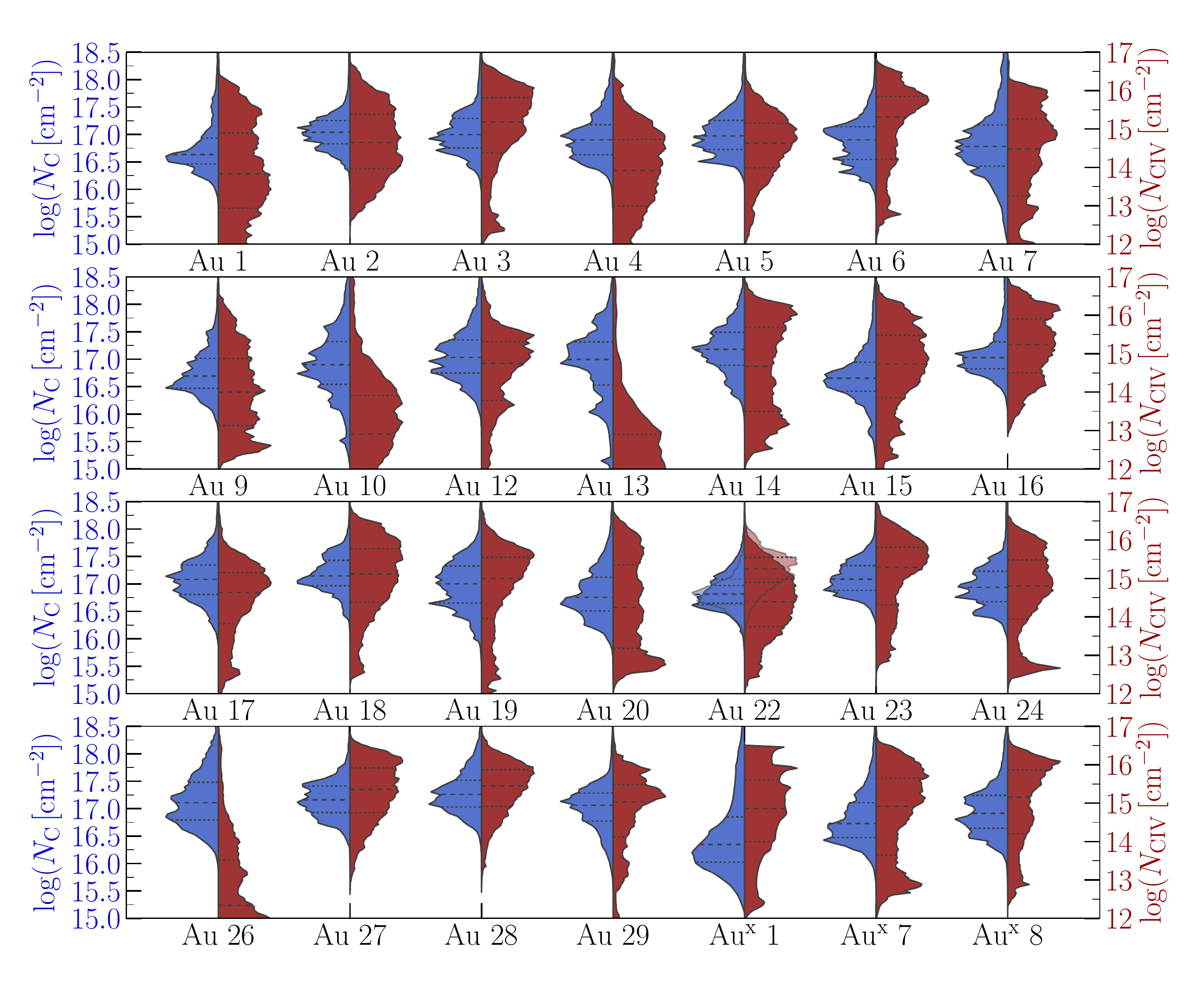}
\caption{A grouped split violin plot representing the column density distributions of carbon (blue, left) and \ion{C}{iv} (red, right) along lines-of-sight (one viewing angle) through the 28 Auriga haloes.}
\label{fig:los-distro-C}
\end{figure*}

\begin{figure*}
\centering
\includegraphics[width=\textwidth]{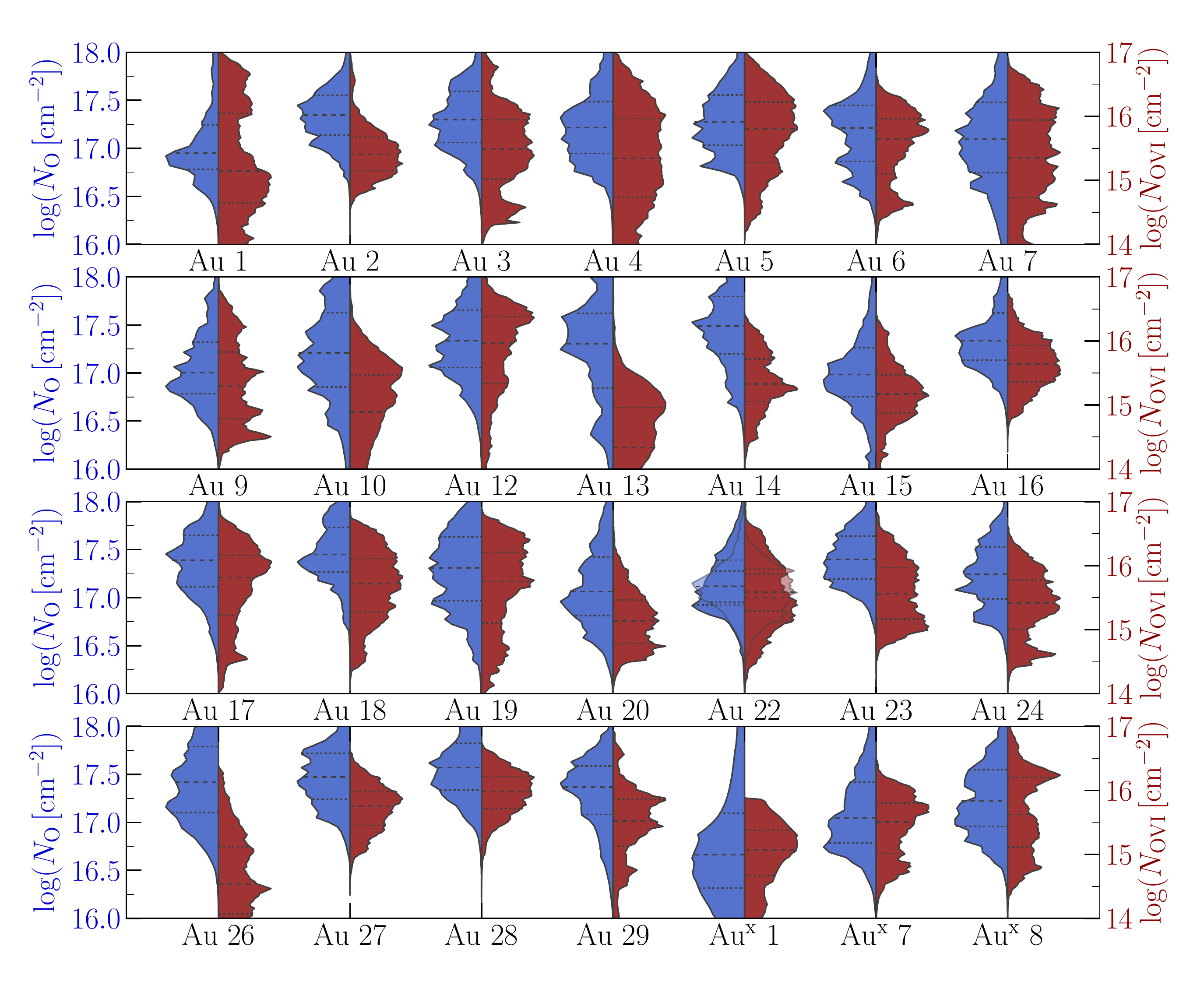}
\caption{A grouped split violin plot representing the column density distributions of oxygen (blue, left) and \ion{O}{vi} (red, right) along lines-of-sight (one viewing angle) through the 28 Auriga haloes.}
\label{fig:los-distro-O}
\end{figure*}

Figures \ref{fig:los-distro-H}--\ref{fig:los-distro-O} show grouped split violin plots representing the column density distributions of hydrogen/\ion{H}{i}, silicon/\ion{Si}{ii}, carbon/\ion{C}{iv}, and oxygen/\ion{O}{vi} along lines-of-sight passing through the Auriga haloes, respectively. Each split violin represents the line-of-sight column density distribution for an element (left) and its corresponding ion (right). The median of the distribution is marked by a dashed horizontal line, while the dotted lines show the interquartile range. Note that the distributions are not commonly normalised and hence should only be used for visual inspection. It is immediately evident that the Auriga haloes display stark differences and variations in their CGM column density distributions. The distributions of each halo span several dex in column densities of metals and hydrogen. The differences in the line-of-sight column density distributions are even more pronounced for the ionic species studied in this work; the column density distributions of \ion{H}{i}, \ion{Si}{ii}, \ion{C}{iv}, and \ion{O}{vi} span $\sim 5$ dex. The wide variation for a given halo is indicative of the complex structure (density, metallicity, temperature) within the CGM. Furthermore, the CGMs of the Auriga haloes convey striking differences: the distributions differ significantly in their shapes and medians, indicating significant diversity in the CGM properties of the simulated sample. Similar diversity in CGM gas properties along different lines-of-sight is reported in other numerical simulations \citep[e.g. ][]{2018MNRAS.477..450N, 2018MNRAS.481..835O, 2019ApJ...873..129P}, and observations of Galactic \citep[e.g. ][]{2017A&A...607A..48R, 2019arXiv190411014W} and extragalactic haloes \citep[e.g. ][]{2004A&A...414...79E,2007A&A...469...61L, 2011Sci...334..948T, 2017ApJ...837..169P}.

\begin{figure*}
\centering
\includegraphics[width=\textwidth]{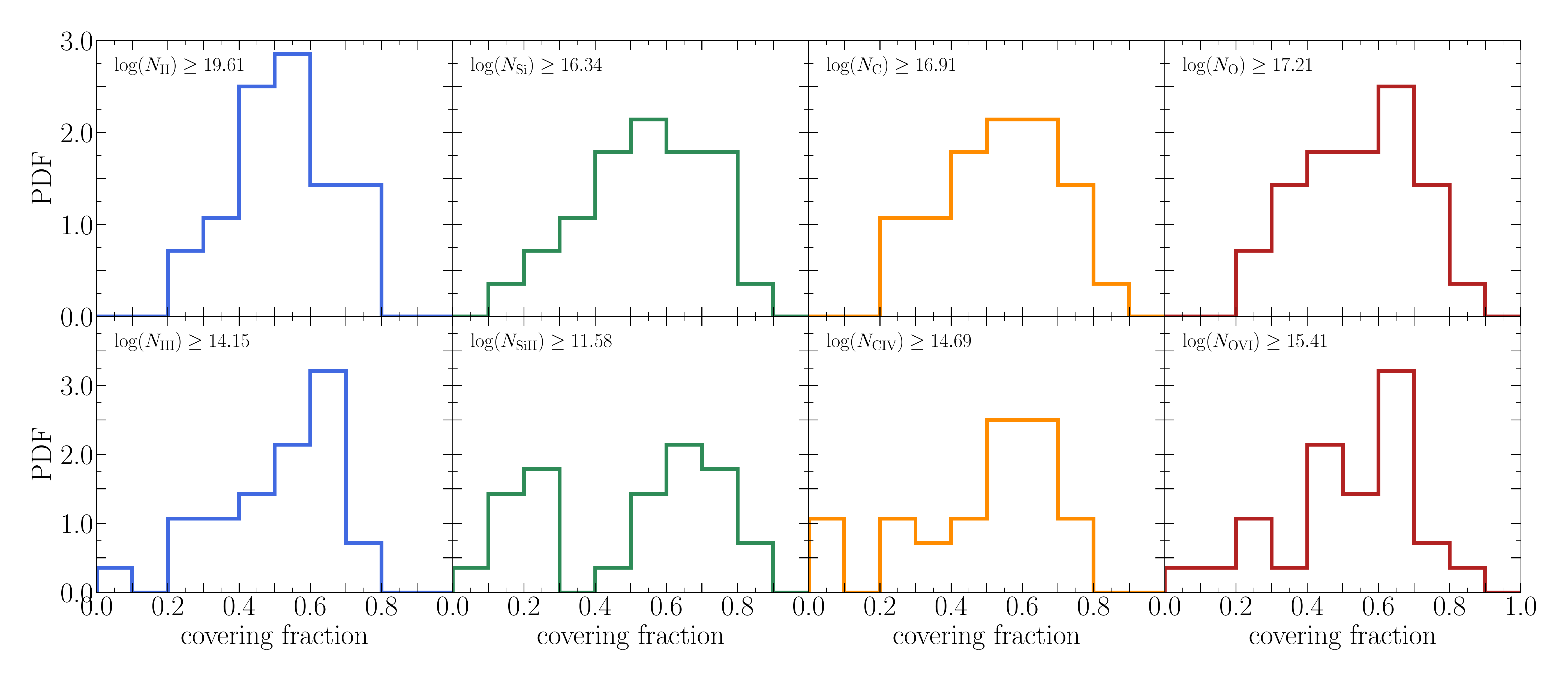}
\caption{The covering fractions of absorbers in the CGM of Auriga galaxies. The top panels show the distributions of hydrogen, silicon, carbon, and oxygen absorbers (left to right) with column densities higher than $10^{19.61}$, $10^{16.34}$, $10^{16.91}$, and $10^{17.21}\,\mathrm{cm}^{-2}$, respectively. The bottom panels show the distributions of \ion{H}{i}, \ion{S}{ii}, \ion{C}{iv}, and \ion{O}{vi} absorbers (left to right) with column densities higher than $10^{14.15}$, $10^{11.58}$, $10^{14.69}$, and $10^{15.41}\,\mathrm{cm}^{-2}$, respectively. Despite the Auriga haloes exhibiting remarkable similarities in their galactic (i.e. halo mass) and environmental properties (i.e. isolated at $z=0$, quiet recent merger histories) their CGM gas shows significant differences.}
\label{fig:CF-hist}
\end{figure*}

Figure \ref{fig:CF-hist} shows the distributions of covering fractions of commonly observed ionic species (bottom panels) and their parent elements (top panels) at representative column densities. The column density thresholds are chosen such that the median cumulative density profile of all the Auriga haloes has a covering fraction of $50\%$. Note that the choice of column density does not affect the qualitative conclusions presented in this work. The covering fractions of all elements and ions vary by about an order of magnitude revealing the vast differences in the CGM structure and content. While some CGMs are rich with dense absorbers --reflected by the high covering fractions of \ion{H}{i}, \ion{Si}{ii}, \ion{C}{iv}, and \ion{O}{vi}, and their parent species-- other CGMs are dominated by comparatively diffuse gas --therefore, exhibiting low covering fractions: $<10\%$ for \ion{H}{i}, \ion{Si}{ii}, \ion{C}{iv}, and \ion{O}{vi}, and $\sim 20\%$ their parent species. The aforementioned diversity in the covering fractions of elements and ions is driven by fundamental differences in the CGM structure and column densities. 

We note that the column densities of \ion{O}{vi} are higher than those inferred from observations \citep[e.g. COS-Haloes][]{2011Sci...334..948T, 2016ApJ...833...54W} and even other simulations with remarkably similar physics models \citep[i.e. Illustris-TNG; ][]{2018MNRAS.477..450N}. The Auriga haloes studied in this work have active AGN (see Figure \ref{fig:sample_properties}) which is not the case in most observational studies. The local AGN radiation field (which is ignored in \citealt{2018MNRAS.477..450N}) drives the high \ion{O}{vi} column densities in our model (equivalently the \ion{Si}{ii} column densities are slightly lower than the observed column densities in L$^\star$ CGMs). As shown in \citet{2018MNRAS.475.1160H}, AGN radiation (modelled in a similar fashion) can be a significant contributor to the ionization in the CGM.

Figures \ref{fig:los-distro-H}--\ref{fig:CF-hist} therefore show that although the subsample of the Auriga haloes used in this study was selected to span a limited range in halo mass, similar isolated environments (see Section \ref{sec:methods-auriga}), and quiet recent merger histories, their CGM reveals significant differences in densities and structure. The CGM of the Auriga haloes bears evidence of their distinct evolution. In the subsequent sections we will highlight some of the major culprits in shaping the CGM of L$^\star$ galaxies.

% --------------------------------------------------------------------------------
% --------------------------------------------------------------------------------
% Stellar mass
% --------------------------------------------------------------------------------
% --------------------------------------------------------------------------------
\subsection{The effects of the host galaxy's stellar mass}
\label{sec:CF-vs-Mstar}
\noindent
The mass--metallicity relationship (MZR) is considered to be one of the most fundamental observed galaxy characteristics: a galaxy's stellar mass correlates (with little intrinsic scatter) with the gas phase metallicity \citep{2004ApJ...613..898T} and the relationship is even tighter when additional variables are considered, such as SFR, size, and gas content \citep[e.g. ][]{2008ApJ...672L.107E,2016A&A...595A..48B}. The MZR (modulo normalization and temporal evolution) holds for a range of galaxy masses \citep[extending from dwarf to L$^\star$ galaxies; e.g. ][]{2006ApJ...647..970L, 2012ApJ...750..120Z}, over a range of redshifts \citep[extending from the local universe up to $z\sim 2$; e.g. ][]{2011ApJ...730..137Z, 2014ApJ...795..165S, 2015ApJ...799..138S}, and at a multitude of scales where recent integral field unit surveys demonstrate the existence of the MZR on kpc scales \citep[i.e. ][]{ 2013A&A...554A..58S, 2017ApJ...844...80B, 2017MNRAS.469.2121S, 2018MNRAS.474.2039E}. The presence and robustness of the MZR highlights the influence of stellar feedback and self-regulation on the galactic gas reservoir, and particularly the importance of the associated star formation and metallicity variation time-scales in defining the MZR \citep[e.g. ][]{2018MNRAS.477L..16T}. 

The presence of an MZR on the galactic scale motivates investigating if such a correlation exists at larger scale viz. the CGM. Stellar feedback is responsible for enriching the ISM via AGB stars, SNe Ia, and SNe II. The enriched gas can then be launched out of the ISM by SN-driven stellar winds or BH-driven winds \citep[see ][]{2005ARA&A..43..769V, 2006MNRAS.373.1265O, 2017MNRAS.468.4170M}. Therefore, given that the processes which are responsible for the MZR on galactic scale also enrich the CGM one may expect the CGM metal content to positively scale with stellar mass. Note that there need not be an MZR in the CGM; unlike the ISM there is no local process regulating the gas content and metal injection in the CGM. Knowing that metal injection in the CGM is driven by feedback processes which depend on ISM properties (i.e. SFR, AGN activity) one may infer that the integrated metal injection into the CGM (i.e. metal mass, or metal covering fraction) scales with galaxy stellar mass which does not necessarily imply an increase in metallicity. \citet{2006MNRAS.371.1125S} demonstrated that a positive scaling between the CGM metal mass and the galaxy stellar mass is a natural outcome of SN feedback.

\begin{figure*}
\centering
\includegraphics[width=\textwidth]{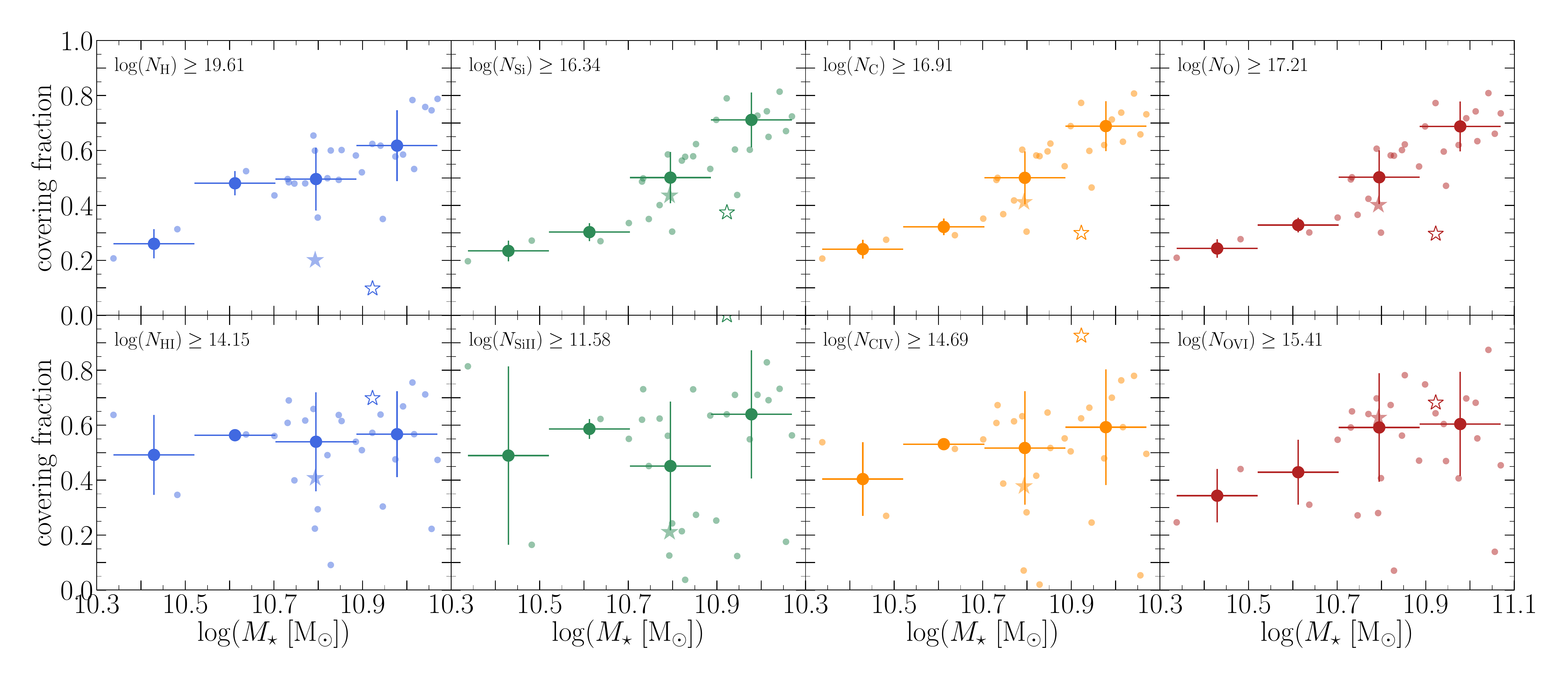}
\caption{The effect of the host galaxy's stellar mass on the CGM gas. The small circles and diamonds show the covering fractions for individual haloes. The large circles represent the median covering fractions in stellar mass bins while the error bars show the bin width and $1\sigma$ variations in covering fraction. The filled and empty stars represent Au 22 and the re-run of Au 22 with no AGN feedback (i.e. Au 22-NOAGN), respectively. The covering fractions of hydrogen, silicon, carbon, and oxygen (top panels) correlate positively with the host galaxy's stellar mass. As the galaxy's mass increases (via star formation or mergers) the CGM is enriched with more metals leading to an enhanced covering fraction. It is worth noting that such an enhancement does not necessarily imply an enhancement in the covering fractions of commonly observed ions as the CGM ionization is dependent on gas density, temperature, and the radiation field. In fact, the covering fractions of \ion{H}{i}, \ion{Si}{ii}, \ion{C}{iv}, and \ion{O}{vi}  (bottom panels) show no correlation with stellar mass.}
\label{fig:CF-vs-Mstar}
\end{figure*}

Figure \ref{fig:CF-vs-Mstar} shows the dependence of covering fraction for ionic species (bottom panels) and their parent elements (top panels) in the CGM on the host galaxy's stellar mass. The positive correlation of the covering fractions of hydrogen, carbon, and oxygen with stellar mass is suggestive of the effect of stellar feedback and galactic outflows on enriching the CGM with metals which correspond to galaxy growth. As the galaxy's mass increases the CGM is increasingly enriched with metals (due to outflows, accretion) which manifests in the increase in the column densities and covering fractions of metals with stellar mass shown in Figure \ref{fig:CF-vs-Mstar}. Unlike the elements, the corresponding ionic species do not show a dependence on stellar mass. The abundance of a given ionic species is driven by the density and temperature structure of the gas as well as the ionizing field. The density and temperature of CGM gas do not immediately scale with stellar mass. Additionally, the incident radiation field (UVB and AGN) is independent of stellar mass in the limited mass range studied in this work. Note that the only ion which shows a possible shallow correlation with stellar mass is \ion{O}{vi} which is thought to be driven by the halo's virial temperature \citep[e.g. ][]{2017MNRAS.465.2966S}. To properly demonstrate the dependence of ionic species on stellar/halo mass one would require a wider range of mass than we have achieved in this work \citep[e.g. ][]{2016MNRAS.460.2157O, 2018ApJ...864..132B}.

The positive correlation of oxygen covering fraction with stellar mass is broadly consistent with other theoretical work. For example, \citet{2018A&A...609A..66M} demonstrated a positive correlation between the mass of oxygen in the CGM and stellar mass over $\sim 2$ decades in stellar mass. A similar dependence of the column densities of oxygen and \ion{O}{vi} on halo mass were also reported by \citet{2016MNRAS.460.2157O}.  Additionally, the results presented in this work agree with the predictions of SN feedback in \citet{2006MNRAS.371.1125S}. Although the correlation of the CGM's oxygen content with the galaxy's stellar mass may seem at odds with current observations \citep[i.e. ][]{2011Sci...334..948T}, computing oxygen mass in observations suffers from various assumptions involved in the ionization corrections. 

% --------------------------------------------------------------------------------
% --------------------------------------------------------------------------------
% Galaxy morphology
% --------------------------------------------------------------------------------
% --------------------------------------------------------------------------------

\begin{figure*}
\centering
\includegraphics[width=\textwidth]{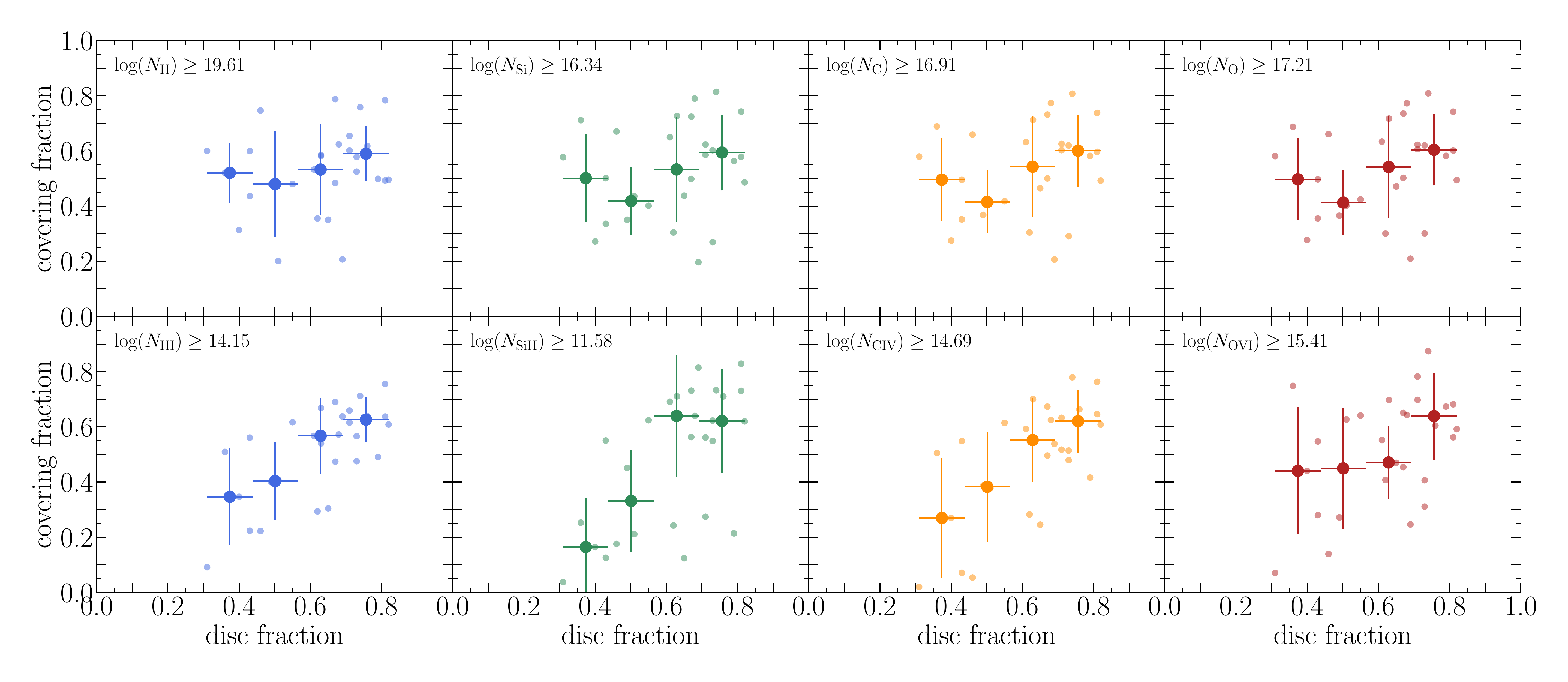}
\caption{The covering fractions of commonly observed ions, \ion{H}{i}, \ion{Si}{ii}, \ion{C}{iv}, and \ion{O}{vi} (bottom panels), and their parent species (top panels) along lines of sight parallel to the galaxy's disc correlate positively with the disc fraction. The small circles show the covering fractions in individual haloes while the large solid circles show the median covering fraction in $D/T$ bins. Outflows driven by stellar feedback and AGN transport gas into the CGM with densities which can sustain low ionization species. A disc-dominated morphology facilitates the ability for clumpy outflows to travel into the CGM without being destroyed near the galaxy. This yields a higher abundance of clumpy clouds in the CGM outside the disc plane therefore increasing the covering fractions shown here.}
\label{fig:CF-vs-DT}
\end{figure*}

\subsection{The effects of galaxy morphology}
\label{sec:CF-vs-DT}
\noindent
Section \ref{sec:CF-vs-Mstar} demonstrates the effect of galaxy growth (i.e. mergers, accretion, star formation) on the CGM metal content where the metals created by stellar feedback are transported into the CGM by galactic outflows which are driven by SNe II or AGN. The physical properties of the outflow depend on the environments from which they are launched and the medium through which they propagate. For example, a wind launched in a dense medium will shock and halt its flow well before reaching the CGM. Therefore, a wind launched in a thin disc will likely escape the disc into a low-density environment in which it can retain its clumpy nature which may sustain low ionization species. Although the decoupling of winds in the simulation reduces the aforementioned effect, the density profile of the medium through which the wind propagates in a disc-like galaxy exhibits a sharper gradient than its counterpart in bulge-dominated galaxies. Therefore, one may still expect to see an effect of galaxy morphology on the ionization of the CGM.

Figure \ref{fig:CF-vs-DT} shows the covering fractions of hydrogen, silicon, carbon, oxygen, \ion{H}{i}, \ion{Si}{ii}, \ion{C}{iv}, and \ion{O}{vi} for different galaxy disc fractions. The disc fraction (i.e. $\mathrm{D/T}$) is calculated following the assumption of \citet[][]{2003ApJ...591..499A} that the bulge component has zero net rotation  hence the bulge mass is twice the mass of the counter rotating component while the disc accounts for everything else\footnote{Note that the results presented in this work are independent of the formalism by which the disc fraction is calculated: i.e. using disc fractions inferred from Gaussian mixture models of dynamical variables shows similar results.}. The covering fractions of \ion{H}{i}, \ion{Si}{ii} and \ion{C}{iv} positively correlate with the disc fraction whilst \ion{O}{vi} (typically tracing hot diffuse gas) shows no correlation with the disc fraction. On the other hand, there appears to be no correlation between the covering fractions of hydrogen, silicon, carbon, or oxygen and the disc fraction which suggests that outflows originating in galaxies with higher disc fraction are more likely to sustain a dense nature (e.g. filamentary, clumpy) where low ionization species can survive.
Gas phase projections reveal visual evidence of the correlation between the filamentary and clumpy structure of the CGM (above and below the disc plane) and the disc fraction. Figure \ref{fig:projection} highlights the vivid filamentary/clumpy structure in the CGM of Au 23 ($\mathrm{D/T} = 0.63$) while the structure in the CGM of Au 26 ($D/T = 0.46$) is deficient. Succinctly, outflows originating in disc-dominated galaxies remain filamentary (see left-hand panels in Figure \ref{fig:projection}) and can host low ionization species after they reach the CGM. Therefore, the CGM of disc-dominated galaxies shows a higher covering fraction of low ionization species when compared to bulge-dominated galaxies (all else being similar). Additionally, the enhanced structure in the CGM along the disc angular momentum axis is consistent with observations \citep[e.g. ][]{2011ApJ...743...10B, 2012ApJ...760L...7K}. For example, the observed strong azimuthal dependence of \ion{Mg}{ii} absorption in inclined disc-dominated galaxies \citep[i.e. ][]{2011ApJ...743...10B, 2012ApJ...760L...7K} is often attributed to galactic outflows \citep[e.g. ][]{2014ApJ...784..108B}. The filamentary outflow structures in disc-dominated galaxies sustain higher densities and can therefore preferentially host low ionization species. However, the results presented in this work are in tension with the observational results of \citet{2015ApJ...815...22K}. The Auriga galaxies show no significant azimuthal dependence of \ion{O}{vi} absorption. \ion{O}{vi} probes the diffuse ionized gas in the CGM which, broadly speaking, shows a smooth profile. It is worth noting that this tension could be artificially induced by the simulation's resolution in the CGM, or the lack of shielding of AGN radiation by intervening gas. The AGN radiation field intensity is computed at the location of a gas cell only considering it's distance from the source ($J_\mathrm{AGN} \propto r^{-2}$) which could possibly overionize the gas in the disc plane which would normally experience the highest levels of shielding.

% --------------------------------------------------------------------------------
% --------------------------------------------------------------------------------
% Star formation
% --------------------------------------------------------------------------------
% --------------------------------------------------------------------------------
\subsection{The effects of recent star formation}
\noindent
Section \ref{sec:CF-vs-Mstar} stresses the effects of galaxy growth on the covering fraction of hydrogen and metals in the CGM. A galaxy's stellar mass reflects its integrated history of star formation, accretion, and mergers. In this section we investigate the effect of recent star formation on the extended CGM. 

\begin{figure*}
\centering
\includegraphics[width=\textwidth]{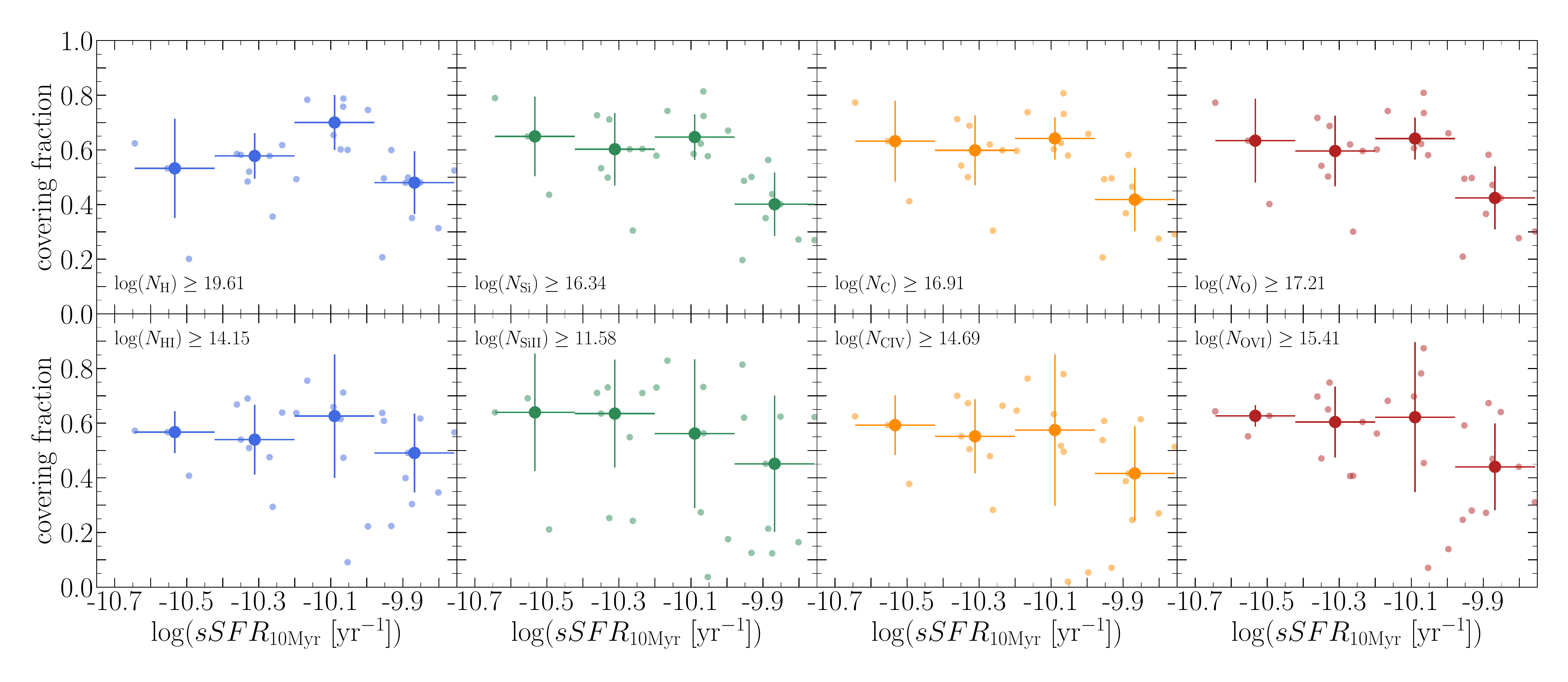}
\caption{The covering fractions of commonly observed ions, \ion{H}{i}, \ion{Si}{ii}, \ion{C}{iv}, and \ion{O}{vi} (bottom panels), and their parent species (top panels) along lines of sight parallel to the galaxy's disc show no correlation the $10$ Myr averaged specific star formation rate. The small circles show the covering fractions in individual haloes while the large solid circles show the median covering fraction in $SFR$ bins.}
\label{fig:CF-vs-SFR}
\end{figure*}

Star formation and the associated stellar feedback (i.e. SN feedback) are main drivers of strong galactic outflows \citep[e.g. ][]{2005ApJ...621..227M, 2005ApJS..160..115R, 2009ApJ...697.2030S} which are often thought to contribute to the observed multiphase, metal-rich CGM. In fact, a correlation between CGM properties and star formation has been reported by various studies \citep[e.g. ][]{2011Sci...334..948T, 2017ApJ...837..169P}. Additionally, studies investigating the CGM of starburst galaxies \citep[e.g. ][]{2013ApJ...768...18B, 2017ApJ...846..151H} suggest that the effect of starburst-driven winds (within $\sim200$ Myr) can be observed out to $\sim 200$ kpc. Although none of the Auriga galaxies studied here are considered starburst galaxies, we explore the effect of recent star formation of the CGM. Figure \ref{fig:CF-vs-SFR} shows the dependence of the covering fractions of \ion{H}{i}, \ion{Si}{ii}, \ion{C}{iv}, and \ion{O}{vi}, and their parent species on the $10$ Myr-averaged specific star formation rate (sSFR). We use the $sSFR$ to account for the correlations shown in Section \ref{sec:CF-vs-Mstar}. The covering fractions of the elements and ions examined in this work show no dependence on the recent star formation history of the galaxies. The lack of correlation between the covering fractions and $sSFR$ also holds when using the $500$ Myr averaged $sSFR$.

% --------------------------------------------------------------------------------
% --------------------------------------------------------------------------------
% AGN radiation et feedback
% --------------------------------------------------------------------------------
% --------------------------------------------------------------------------------
\subsection{The effect of the central engine}
\noindent
Supermassive BHs residing at the centres of galaxies are thought to play a major role in galaxy evolution. Variations in galaxy and halo properties are often attributed to feedback from AGN in the form of radio jets \citep{2012ARA&A..50..455F}, AGN-driven winds \citep[e.g. ][]{2005ApJ...632..751R, 2013MNRAS.436.3031V, 2017ApJ...839..120W}, or a strong radiation field originating from accretion discs surrounding central supermassive BHs \citep[e.g. ][]{2017MNRAS.471.1026S, 2018MNRAS.tmp.2172O, 2018MNRAS.474.4740O}. \citet{2018MNRAS.478.3890B} show that AGN hosts show an enhancement in low ionization species (e.g. \ion{H}{i}, \ion{Si}{ii}, \ion{Si}{iii}) in their CGM at large radii, hence suggesting that the enhancement might be caused by accretion or the remnants of a previous activity (e.g. AGN episode, starburst). In this section we investigate the effects of AGN feedback on the CGM of L$^\star$ galaxies.

%
%%
%%% AGN radiation
\subsubsection{The AGN radiation field}
\noindent
The accretion disc surrounding central BHs provides a strong radiation field which can have a significant effect on gas ionization. \citet{2018MNRAS.474.4740O} demonstrate that an AGN can keep the CGM ionized for time-scales which significantly exceed the AGN lifetime leading to proximity fossil zones. Additionally, \citet{2018MNRAS.tmp.2172O, 2018MNRAS.474.4740O} used hydrodynamical simulations with non-equilibrium ionization to demonstrate the importance of AGN radiation when reproducing the observed \ion{O}{vi} covering fractions. Moreover, \citet{2018MNRAS.475.1160H}, using an AGN model identical to the one implemented in the work presented here, showed the strong effect of the AGN on the ionization state of the CGM.

\begin{figure*}
\centering
\includegraphics[width=\textwidth]{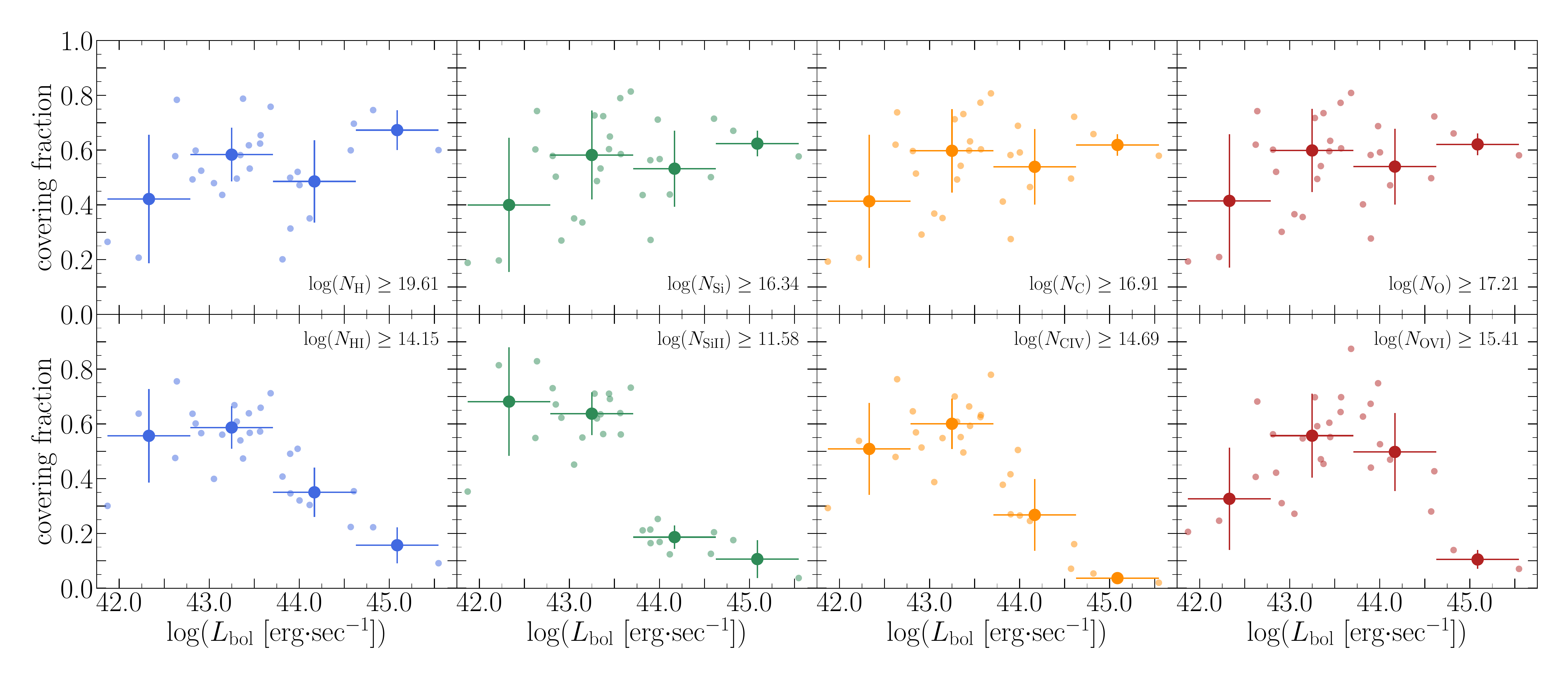}
\caption{The effect of the AGN on the CGM ionization is evident in the Auriga model. The covering fractions of \ion{H}{i}, \ion{Si}{ii}, \ion{C}{iv}, and \ion{O}{vi} (bottom panels) and their parent species (top panels) are shown as a function of the AGN bolometric luminosity. The small circles show the covering fractions of individual Auriga haloes while the large circles show the median covering fractions in bins of $L_\mathrm{bol}$ and the error bars show bin sizes and $1\sigma$ variations in binned covering fractions.  The AGN's radiation field is a significant source of ionization which drives a decrease in the covering fractions of commonly observed ions.}
\label{fig:AGN_CF-vs-Lbol}
\end{figure*}

Figure \ref{fig:AGN_CF-vs-Lbol} shows the covering fractions of hydrogen, carbon, oxygen, \ion{H}{i}, \ion{C}{iv}, and \ion{O}{vi} as a function of the AGN bolometric luminosity in the Auriga haloes. As the AGN luminosity increases the ionization of the CGM increases due to the aggressive AGN radiation field. The increase in CGM ionization manifests as a decrease in the covering fractions of the ionic species studied in this work. For AGN luminosities $L_\mathrm{bol}\gtrsim 10^{44.7} \, \mathrm{erg s}^{-1}$, the CGM densities become insufficient for any of the ionization species studied here (i.e. \ion{H}{i}, \ion{Si}{ii}, \ion{C}{iv}, and even \ion{O}{vi}) to survive in significant quantities. This effect is more drastic for the low ionic metal species (i.e. \ion{Si}{ii}) where the covering fractions decline drastically at even lower AGN luminosities ($L_\mathrm{bol}\gtrsim 10^{43.7} \, \mathrm{erg s}^{-1}$).

We caution the reader that the ionization calculations were performed in post-processing assuming ionization equilibrium which poses a limitation for the analysis presented here. Various works have highlighted the importance of considering non-equilibrium effects in the CGM especially in the presence of a stochastically varying AGN \citep[i.e. ][]{2013MNRAS.434.1063O, 2018MNRAS.474.4740O, 2018MNRAS.481..835O}. \citet{2017MNRAS.471.1026S} stress that metals in the CGM may be out of ionization equilibrium in the presence of a fluctuating radiation field (i.e. AGN) even at large radii (i.e. $2\times R_\mathrm{vir}$). Non-equilibrium ionization effects are particularly important when the times-cales on which the AGN varies are comparable to the recombination time-scales of the ions of interest $\sim 20 \, \mathrm{Myr}$ (i.e. \ion{H}{i}, \ion{Si}{ii}, \ion{C}{iv}, \ion{O}{vi}). In the simulations we present, the AGN variability time-scales are of order $45\, \mathrm{Myr}$ (calculated using the power spectral density) which well exceeds the recombination timescales. Moreover, there is no evidence for a proximity zone fossil in the simulations we present which is expected with the long AGN variation timesscales \citep[i.e. ][]{2018MNRAS.474.4740O}. Therefore, although the assumption of ionization equilibrium is a potential limitation of this work non-equilibrium effects appear to be negligible in the simulations we present.

%
%%
%%% AGN feedback
\subsubsection{AGN-driven feedback}
\noindent
AGN feedback is often introduced in simulations of galaxy evolution to regulate star formation, quench massive galaxies, and reproduce the high-mass tail of the galaxy mass function and the cosmic star formation history \citep[e.g. ][]{2006MNRAS.370..645B, 2013MNRAS.428.2966P, 2014MNRAS.442.2751T}. However, AGN feedback has also been shown to have a noticeable impact on the CGM by effectively launching gas into and outside of the CGM \citep{2017MNRAS.467..179G} and creating a hot galactic halo by thermally heating the CGM \citep{2015ApJ...804...72B}.

The physical model used in this work includes AGN feedback in both quasar and radio mode. The quasar mode isotropically deposits thermal energy into the BH's neighbouring gas cells, while the radio mode exerts mechanical feedback by isotropically and gently inflating hot buoyant bubbles in the halo. For further description of the AGN feedback model see Section \ref{sec:methods}, \citet{2013MNRAS.436.3031V}, \citet{2014MNRAS.437.1750M}, and \cite{2017MNRAS.467..179G}. In order to better understand the effect of AGN feedback on the CGM we use a re-run of Au 22 which excludes feedback from the AGN \citep[hereafter Au 22-NOAGN;  ][]{2017MNRAS.467..179G}. The re-run simulates Au 22 with no changes until redshift $z=1$ after which AGN feedback is eliminated while accretion on to the BH continues. Therefore, differences between Au 22 and Au 22-NOAGN can be attributed to AGN feedback. 

AGN feedback reduces the central star formation and therefore produces a smaller bulge component. In the absence of AGN feedback, the galaxy grows a more prominent bulge component thus decreasing the disc fraction \citep[i.e. ][]{2017MNRAS.467..179G}. Globally, the galaxy increases its stellar mass from $10^{10.79}$ to $10^{10.93}\, \mathrm{M}_\odot$. The increased star formation is driven by gas condensation from the CGM onto the central galaxy. In the presence of AGN feedback, the halo gas is too hot and unable to cool and form stars. As a result, the halo is hotter and more populated with gas. At $z=0$, the bolometric AGN luminosities are $10^{43.81}$ and $10^{40.05} \, \mathrm{erg s^{-1}}$ for Au 22 and Au 22-NOAGN\footnote{Note that although AGN feedback is eliminated, accretion onto the black hole is permitted.} respectively. For a comprehensive discussion of the effect of the AGN on the galactic properties (briefly discussed above) see section 4.3 and figure 17 in \citet{2017MNRAS.467..179G}.

Returning to Figures \ref{fig:los-distro-H}--\ref{fig:los-distro-O}, which show the column density distributions of lines-of-sight through the Auriga haloes, the column density distributions of lines-of-sight through Au 22-NOAGN are shown as the transparent distributions. The halo with no AGN feedback (Au 22-NOAGN) shows significant differences in its hydrogen column density distribution compared to its AGN counterpart (Au 22) where Au 22 shows a shift towards higher hydrogen column densities. AGN feedback stifles accretion and condensation from the CGM onto the galaxy therefore yielding a richer and hotter gas halo and lower galaxy stellar masses. Regardless of the lower hydrogen column densities in Au 22-NOAGN, the halo exhibits higher \ion{H}{i} column densities owing to the lower halo AGN luminosity. Unlike hydrogen, the metal column density distributions of Au 22 and Au 22-NOAGN are remarkably similar. Nonetheless, the two haloes differ strongly in their ionization: Au 22-NOAGN has its \ion{Si}{ii} and \ion{C}{iv} distributions shifted to significantly higher column densities while the \ion{O}{vi} column density distributions remain unchanged. The increased abundances of \ion{Si}{ii} and \ion{C}{iv} in Au 22-NOAGN is driven by the lower AGN luminosity (note that the radiation field driven by the AGN aggressively ionizes the CGM). However, the presence of \ion{O}{vi} in the halo is primarily driven by the halo virial temperature at low AGN bolometric luminosities where ionization is independent of the AGN. Knowing that Au 22 and Au 22-NOAGN reside in haloes with similar halo mass, their \ion{O}{vi} column density distributions are similar. 

Comparing Au 22 and Au 22-AGN provides a constraint on the effect of AGN feedback on the properties of a halo. While understanding the effects of AGN feedback through direct comparison of Au 22 and Au 22-NOAGN is key, a further comparison is warranted in the framework of the analysis presented here. Figure \ref{fig:CF-vs-Mstar} highlights Au 22 and Au 22-NOAGN by the filled and clear star symbols. Although Au 22-NOAGN's CGM displays similar metal column density distributions to its AGN counterpart, it remains severely underpopulated with metals compared to haloes with similar stellar mass. Au 22-NOAGN is displaced by $\sim 3\sigma$ from the correlation between metal covering fraction and stellar mass. Moreover, the CGM of Au 22-NOAGN is severely underpopulated with hydrogen. The stark differences between the CGM of Au 22-NOAGN and the CGMs of haloes with similar stellar mass (i.e. stellar mass matched control) emphasize the key role AGN feedback plays in controlling accretion onto the galaxy, ionizing the CGM, and populating the CGM with metals.

% --------------------------------------------------------------------------------
% --------------------------------------------------------------------------------
% Merger histories
% --------------------------------------------------------------------------------
% --------------------------------------------------------------------------------
\subsection{The effects of galaxy assembly}
\noindent
Galaxy mergers are thought to play a key role in galaxy assembly and evolution where galaxies grow their stellar masses \citep{1993MNRAS.262..627L}, alter their morphologies \citep[e.g. ][]{1987ApJ...312....1H, 1996ApJ...471..115B, 2008MNRAS.391.1137L, 2014MNRAS.445.1157C, 2016MNRAS.461.2589P, 2018MNRAS.480.1715P}, and enhance their star formation rates \citep[e.g. ][]{2006MNRAS.373.1013C, 2008MNRAS.384..386C, 2008A&A...492...31D, 2011MNRAS.412..591P, 2013MNRAS.430.3128E, 2015MNRAS.454.1742K, 2015MNRAS.448.1107M, 2016MNRAS.462.2418S, 2018MNRAS.479.3381B} and AGN activity \citep[e.g. ][]{2005Natur.433..604D, 2010MNRAS.407.1529H, 2011MNRAS.418.2043E, 2014MNRAS.441.1297S, 2017MNRAS.464.3882W}. Additionally, recent work has shown that mergers increase the atomic and molecular gas content in galaxies \citep[e.g. ][]{2018MNRAS.478.3447E,  2018ApJ...868..132P, 2018MNRAS.476.2591V} which might be expected to manifest as changes in the CGM as well. Not only do galaxy mergers impact galaxy stellar populations, mergers influence the gas properties in the CGM. Merger-induced shocks and feedback can heat the CGM gas \citep[e.g. ][]{2004ApJ...607L..87C, 2006ApJ...643..692C, 2009MNRAS.397..190S}, while outflows driven by the associated AGN activity \citep[e.g. ][]{2005ApJ...632..751R, 2013ApJ...776...27V, 2017ApJ...839..120W} and enhanced star formation \citep[e.g. ][]{2005ApJ...621..227M, 2005ApJS..160..115R, 2017MNRAS.465.1682H} populate the CGM with metals and give rise to a multiphase medium. For example, \citet{2018MNRAS.475.1160H} demonstrated the impact of gas-rich major mergers on the CGM showing a case study of a merger occurring at $z \sim 0.66$. The authors report a significant long lasting increase in the CGM metallicity and metal content, as well as an increase in ionization. Besides, \citet{2017MNRAS.467..179G} demonstrated, using the Auriga galaxies, that galaxy mergers play an important role in changing the halo angular momentum, hence influencing the size of the galactic disc (quiescent mergers grow the galactic disc while violent mergers destroy the disc and induce AGN activity which suppresses outer disc formation).

Understanding the extent of the effects of galaxy mergers on the CGM requires a comprehensive study: for example, one must first understand the impact of basic merger properties such as the merger mass ratio, merger geometry, and galaxy gas fractions before properly comprehending the effect of complex integrated merger histories. Nonetheless, one can still evaluate the relative importance of mergers (compared to other processes) in building and shaping the CGM of galaxies in the local universe by investigating the strength of the correlation between a galaxy's CGM properties and merger history. Note that although we explicitly remove interacting systems and recent post-mergers from our analysis (Section \ref{sec:methods-auriga}), we can still investigate the effect of the integrated merger history (\textit{not recent mergers}) on the CGM properties. The most recent major merger (mass ratio $> 1:3$ ) in our sample happened at $t_\mathrm{lookback} \sim 2.3 \, \mathrm{Gyr}$, and the most recent minor merger (mass ratio $>1:10$) happened at $t_\mathrm{lookback}\sim 2.1 \, \mathrm{Gyr}$. In this work we quantify the merger history by computing the stellar mass fraction accreted since $z=2$ (ex-situ stellar mass fraction) as described by \citet{2017MNRAS.472.3722G} and \citet{2019MNRAS.485.2589M}. Although this metric has no way of distinguishing between the details of the merger histories it gives a broad brush understanding on the importance of mergers in shaping the CGM of galaxies.

\begin{figure*}
\centering
\includegraphics[width=\textwidth]{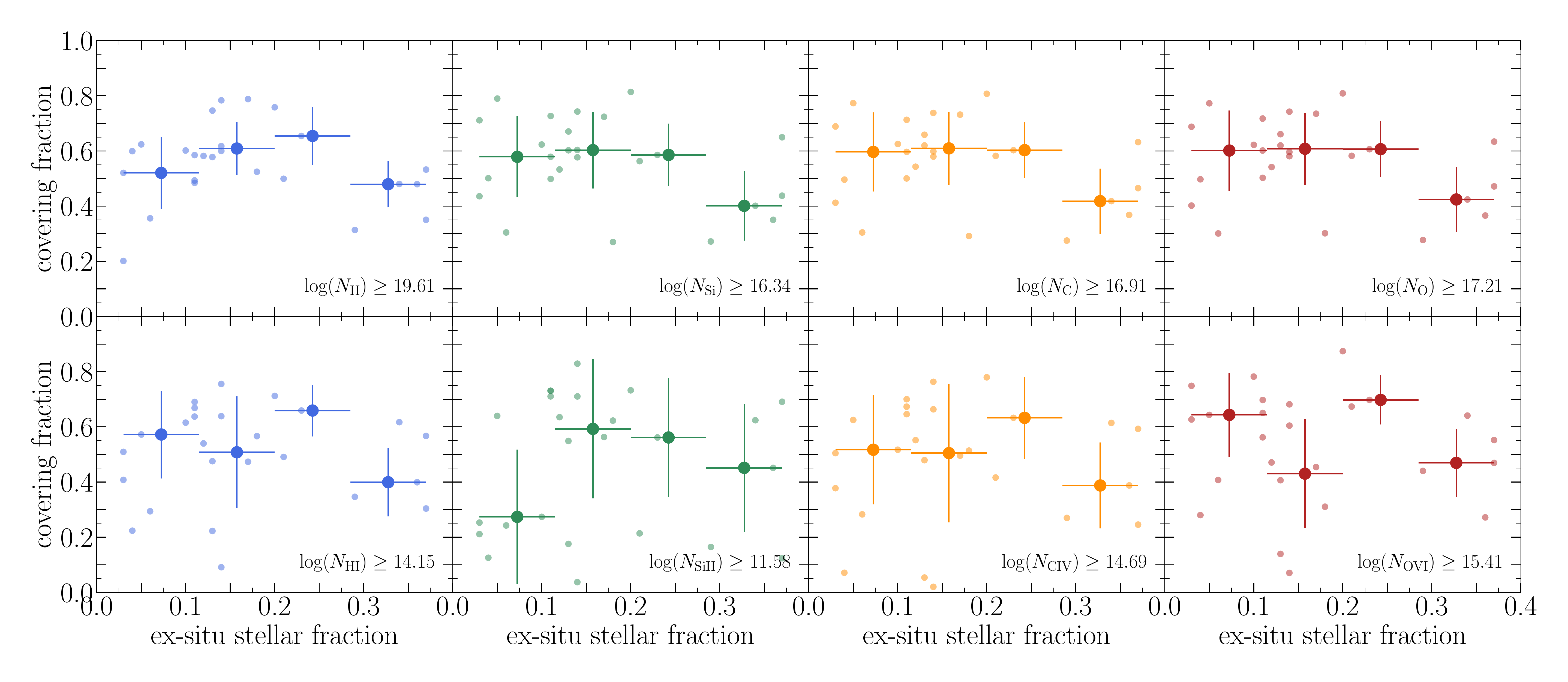}
\caption{The effect of the merger history on the CGM properties. The covering fractions of \ion{H}{i}, \ion{Si}{ii}, \ion{C}{iv}, and \ion{O}{vi} (bottom panels) and their parent species (top panels) are shown as a function of the ex-situ stellar mass fraction. The small circles represent the covering fractions of individual haloes, while the large solid circles show the median covering fractions in bins of ex-situ stellar mass fraction. The error bars show the bin size and $1\sigma$ variations in the binned covering fractions. The absence of correlation between the CGM metal content or ionization and the merger histories suggests that the galaxy merger history is not a major determinant of CGM properties in the absence of recent mergers. Other processes are at least as impactful as galaxy mergers in shaping the CGM of $z=0$ L$^\star$ galaxies.}
\label{fig:CF-vs-facc}
\end{figure*}

Figure \ref{fig:CF-vs-facc} shows the dependence of the covering fraction of hydrogen, silicon, carbon, oxygen, \ion{H}{i}, \ion{Si}{ii}, \ion{C}{iv}, and \ion{O}{vi} on the galaxies` merger history (ex-situ stellar mass fraction). The lack of correlation between CGM properties (covering fraction) and the merger history (ex-situ stellar mass fraction) hints to other processes being at least as dominant as a galaxy's merger history (in the absence of a recent merger) in shaping the CGM. It is, however, worth noting that the galaxies that show the lowest covering fractions have not had a merger more major than $1:10$ since $z=2$. Nonetheless, not all galaxies with quiet merger histories exhibit low metal covering fractions in their CGM.

\section{Discussion}
\label{sec:discussion}
\noindent
Observational surveys targeting the CGM are often dependent on the fortuitous coincidence of bright UV quasars and foreground galaxies \citep[e.g. ][]{2013ApJ...777...59T,2014ApJ...796..136B,2016ApJ...833..259B,2017ApJ...846..151H, 2017ApJ...837..169P, 2018MNRAS.478.3890B}. Inferring CGM properties through absorption features provides a sensitive and powerful tool for CGM studies. Current observational surveys have painted a solid foundation to our understanding of the CGM and its relation to galaxy properties in a statistically meaningful way  \citep[e.g. ][]{2010ApJ...708..868C, 2011ApJ...740...91P, 2013ApJ...777...59T, 2014MNRAS.445.2061L, 2015ApJ...804...79L}. Nonetheless, these observational studies suffer from two major limitations: (1) small samples of galaxy--quasar pairs, and (2) small (often one) number of sight-lines through each galaxy. Therefore, studying the CGM of single galaxies in absorption and understanding its diversity and structure is not currently possible. Consequently, observational surveys often resort to two fundamental assumptions: (1) the CGM of a halo has small azimuthal variations and therefore a single sight-line at a given impact parameter is representative of the CGM at the observed impact parameter, and (2) a well-controlled (often controlled for stellar mass) sample of galaxies has little variations between their CGMs and therefore the observed sight-lines can be used to reproduce a representative median CGM profile defined by the control parameters.

Characterizing the diversity in the CGM within a controlled sample requires an unfeasible number of lines-of-sight and galaxies. However, various observational studies tackled the variations within the CGM of single haloes. Studies using lensed quasars demonstrate a large dispersion in observed absorption properties (line strength, and equivalent widths) of low and intermediate ionization species (unlike high ions) over kpc scales \citep[e.g. ][]{2004A&A...414...79E,2007A&A...469...61L,2017ApJ...851...88K}. The lensed quasar findings are consistent with those of studies using multiquasar systems which report strong variations in column densities and kinematics within the CGM of single haloes \citep[e.g. ][]{2016ApJ...826...50B}. Additionally, the Galactic CGM has been studied with a plethora of lines-of-sight stressing the diversity in the CGM \citep[e.g. ][]{2017ApJ...834..155M, 2017A&A...607A..48R, 2019arXiv190411014W}. Such variations within observed CGMs highlight the complexity of these reservoirs' structure.

% introducing the mock surveys
Section \ref{sec:results-diversity} demonstrates the diversity within a given halo's CGM and within the Auriga sample which poses a major challenge to understanding the CGM with limited sampling. Assuming that the diversity in the Auriga CGMs is representative of the observed universe, we investigate the effects of the limited sampling of the \textit{diverse} CGM in observational studies has on reproducing CGM properties. We create mock observational surveys of the CGM of Auriga galaxies whilst controlling for either stellar mass (within $\pm \delta \log(M_\star) $ centred at $ \log(M_\star) = 10.8$), or AGN luminosity (within $\pm \delta \log (L_\mathrm{bol})$ centred at $\log(L_\mathrm{bol}) =43.0$), hereafter referred to as the controlled sample. For each controlled sample we pool all the lines-of-sight passing through the CGM of the member galaxies. Then we sample, without replacement, $N_\mathrm{los}$ lines-of-sight to produce a mock survey for which we compute covering fractions for the elements and ionic species used in this work. The lines-of-sight are then returned to the controlled sample pool and the sampling process is repeated $10^6$ times. The inferred covering fractions are then compared to the  \textit{true} covering fraction which is computed for a hypothetical halo including all the lines-of-sight in the controlled sample pool. 

\begin{figure*}
\centering
\includegraphics[width=\textwidth]{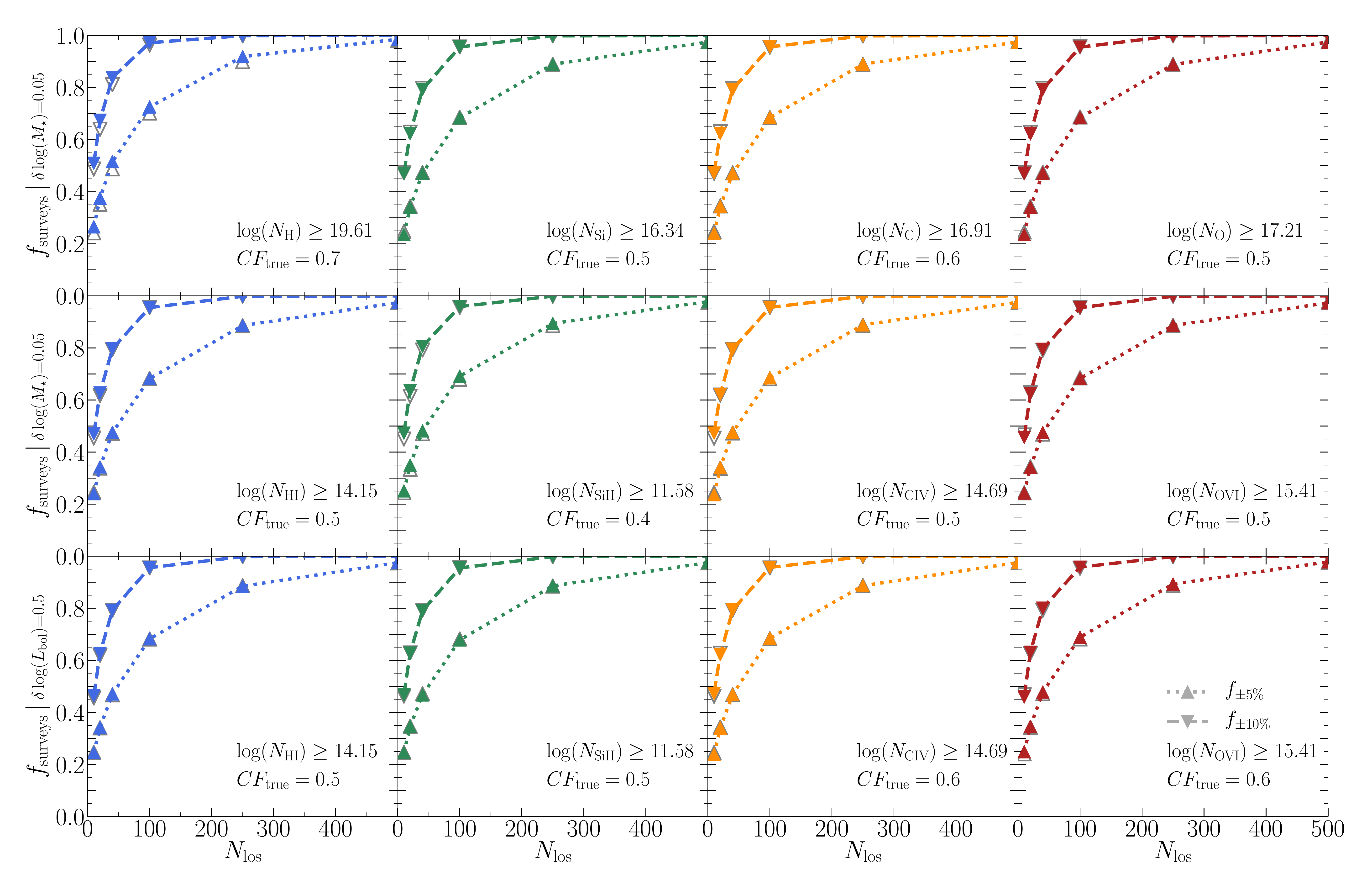}
\caption{Quantifying the effect of limited sampling on measuring the \textit{true} covering fraction of various elements and ionic species. We report the fraction of mock surveys with an inferred covering fraction within $\pm 5\%$ (dotted line/upwards triangles) and $10\%$ (dashed line/downwards triangles) of the \textit{true} covering fraction as a function of the number of lines-of-sight in the survey. The top and middle rows show the results for a stellar mass-matched sample ($\delta \log (M_\star) = 0.05$) while the bottom row shows the effect of $L_\mathrm{bol}$-matching ($\delta \log (L_\mathrm{bol}) = 0.5$). The species whose covering fraction is calculated is shown in the bottom right of each panel. Decreasing $N_\mathrm{los}$ produces a poorer estimate of the \textit{true} covering fraction which is an effect of limited sampling. Consequently, an impractically large number of lines-of-sight is needed to confidently reproduce the underlying CGM properties. The open grey triangles represent a sample which has been matched with $\delta \log (M_\star) = 0.25$ (top and middle rows), or $\delta \log (L_\mathrm{bol}) = 1.5$ (bottom row). The matching criteria has a negligible effect on the probability of estimating the \textit{true} covering fraction which may suggest that the diversity within a single CGM overshadows the diversity within a controlled sample for the Auriga galaxies studied in this work.}
\label{fig:mock-survey}
\end{figure*}

% Nlos 
Figure \ref{fig:mock-survey} shows the fraction of mock surveys which reproduce the \textit{true} covering fraction to within $\pm 5\%$ and $\pm 10\%$ as a function of $N_\mathrm{los}$ for hydrogen, silicon, carbon, and oxygen (top panels), and \ion{H}{i}, \ion{Si}{ii}, \ion{C}{iv}, and \ion{O}{vi} (middle and bottom panels). The top two rows show the results for a stellar mass matched sample ($\delta \log (M_\star) = 0.05$), whilst the bottom row uses a sample which has been matched for $L_\mathrm{bol}$ ($\delta \log (L_\mathrm{bol}) = 0.5$). As the number of lines-of-sight increases, the inferred covering fraction distribution becomes tighter. Consequently, the probability of recovering the \textit{true} covering fraction of the sampled CGM increases. For example, we find that a survey typically needs at least $250$ lines-of-sight in order to constrain the covering fraction to within $5\%$. We note, however, that when deriving the scaling relations shown in this work a tolerance of $\sim 10\%$ is sufficient. Therefore, one may be able to reproduce the results presented above using a lower number of lines-of-sight.

% matching criteria
The open grey triangles in Figure \ref{fig:mock-survey} represent a sample which has been matched with less stringent matching criteria: $\delta \log (M_\star) = 0.25$, and $\delta \log (L_\mathrm{bol}) = 1.5$ (i.e. including all the Auriga galaxies studied in this work). The matching criteria has a negligible effect on the probability of a mock survey reproducing the \textit{true} covering fraction. The negligible effect of the matching criteria suggests that the diversity within a single CGM overshadows the diversity within a controlled sample for the Auriga haloes studied in this work.

The results of this analysis hint that while matching the galaxy sample by stellar mass and AGN luminosity has a negligible effect on reproducing the \textit{true} covering fraction of the sample, an impractically large number of lines-of-sight is needed to reproduce the \textit{true} covering fraction with high confidence. However, owing to the narrow range of matching investigated in this work and the narrow range of galaxy properties spanned by the Auriga haloes studied here, one must proceed with caution when considering the importance of controlled galaxy samples: viz. the independence of the success of mock surveys on the matching criteria may not necessarily imply the unimportance of sample matching. 

One must be cautious when extrapolating the diversity in the Auriga sample into the observed universe. Although the physical model used in this work reproduces many galactic properties \cite[see ][]{2017MNRAS.467..179G} and some observed CGM properties (i.e. azimuthal dependence of low ionization species, dependence of the CGM gas properties on AGN luminosity and stellar mass), the achieved resolution in the CGM may pose a limitation to the analysis we present. Recent studies targeting the CGM introduce refinement schemes to better resolve the underlying structure which contributes to most of the absorption and demonstrate the importance of high resolution in CGM studies \citep[i.e. ][]{2019ApJ...873..129P, 2019MNRAS.483.4040S, 2019MNRAS.482L..85V}. \citet{2019MNRAS.483.4040S} and \citet{2019ApJ...873..129P} show that although the structure of the CGM changes drastically with increasing resolution the integrated quantities show smaller changes; harmoniously, \citet{2019MNRAS.482L..85V} and \citet{2019arXiv190110777P} independently demonstrate the strong dependence of column densities and covering fractions of \ion{H}{i} on resolution. Additionally, \citet{2019MNRAS.483.4040S} demonstrate the effect of the resolved structure at producing and sustaining the cool gas in the CGM.  We also stress that the galaxy sample studied in this work spans a narrow range in halo and stellar mass, and is selected to be isolated at $z=0$, and to have a quiet recent merger history all of which are factors that are worth exploring when comparing to analogous observational samples.

\section{Conclusions}
\label{sec:conclusions}
\noindent
The CGM is a major gas and metal reservoir that plays a key role in galaxy evolution. Whilst the CGM is ubiquitous, its detailed properties are complex and dependent on a plethora of galactic and environmental properties. We use 28 simulated galaxies from the Auriga project, a suite of magneto-hydrodynamical cosmological zoom-in simulations \citep[][]{2017MNRAS.467..179G}. The galaxies are chosen to exhibit no signs of current or recent interactions with a narrow range of halo mass. We examine the diversity within the galaxy sample and individual CGMs, and we investigate correlations with well-understood galaxy properties (e.g. stellar mass, disc fraction, AGN activity, merger history). 
We have demonstrated that: 
\begin{enumerate}
    \item{The CGM of individual haloes shows remarkable diversity in the distribution of column densities for hydrogen, metals, and their associated ionic species studied in this work. The column density distributions (even within a single halo) span $4-5$ decades for the ionic species (\ion{H}{i}, \ion{Si}{ii}, \ion{C}{iv}, and \ion{O}{vi}), and $2-3$ decades for their parent species. Although the Auriga haloes studied in this work were chosen to occupy a narrow range in halo mass, and have relatively quiet merger histories, they show stark diversity in their CGM properties, covering fractions (a decade in covering fraction), and line-of-sight column density distributions (several decades in column density) for commonly observed ions and their parent species. This is indicative of the sensitivity of the CGM to galaxy evolution.}
    \item{Ionic species (e.g. \ion{H}{i}, \ion{Si}{ii}, \ion{C}{iv}, and \ion{O}{vi}) do not obey the same scaling relations as the total O, C, Si, and H (see Figures \ref{fig:CF-vs-Mstar} and \ref{fig:AGN_CF-vs-Lbol}) due to the effects of ionization. Therefore, when interpreting observations it is important to properly account for ionization effects.}
    \item{The CGM gas and metal content (i.e. covering fractions) show a tight correlation with the host galaxy's stellar mass. As galaxies grow, the metals produced in the ISM are transported to the CGM via stellar or AGN winds, therefore increasing the CGM's metal content. Conversely, the CGM ionization is independent of the galaxy's stellar mass as the gas ionization is sensitive to the density and temperature distributions as well as the ionization field.}
    \item{While the CGM gas and metal content is independent of current AGN activity (i.e. $L_\mathrm{bol}$), the CGM ionization is anticorrelated with  $L_\mathrm{bol}$. The AGN's ionization field provides a major source of ionization in the CGM.}
    \item{In addition to being a major driver of the CGM's ionization, the AGN can have a noticeable effect on the CGM's gas and metal content. AGN-driven winds transport metals from the ISM to the CGM, while AGN feedback can suppress accretion onto the galactic disc, thus increasing the CGM's gas (hydrogen) content.}
    \item{The CGM gas and metal content is independent (with large scatter) of the recent ($10-500$ Myr averaged) specific star formation rate.}
    \item{The CGM ionization shows a positive correlation with the galaxy's disc fractions. Outflows originating in disc-dominated galaxies can travel unperturbed into the CGM, thus maintaining its filamentary and dense structure that allows the existence of low ionization species. Nonetheless, the covering fraction of \ion{O}{vi} is independent of disc fraction as it traces the hot diffuse medium.}
    \item{No correlation is present between the CGM properties and the long-term galaxy merger history (i.e. ex-situ stellar mass fraction) which indicates that mergers during the early stages of galaxy evolution are not a dominant sculptor of the $z=0$ CGM in the absence of recent major mergers. We remind the reader that the Auriga galaxies studied in this work were selected to have quiet recent merger histories (no mergers more major than $1:3$ within the past $2$ Gyr).}
    \item{The diversity of the Auriga CGM stresses the need for impractically large numbers of lines-of-sight when studying the CGM. As a result, with $\sim 40$ lines-of-sight, only $\sim 50\%$ of mock surveys would infer a covering fraction within $5\%$ of the \textit{true} covering fraction of a controlled galaxy sample.}
\end{enumerate}

This work presented a sample of $z=0$ Milky Way-mass (L$^\star$) galaxies simulated in a realistic cosmological environment, and demonstrated the stark diversity within an individual CGM and between the CGMs of the sample. Additionally, we highlight major physical and evolutionary processes that mould the CGM, its gas and metal content, ionization, and structure. The diversity of the Auriga CGM accentuates a fundamental challenge for observational studies.
\section*{Acknowledgements}
The authors thank the anonymous referee for their helpful comments which improved the presentation of this work. MHH acknowledges the receipt of a Vanier Canada Graduate Scholarship. SLE acknowledges the receipt of an NSERC Discovery Grant. MS acknowledges support by the European Research Council under ERC-CoG grant CRAGSMAN-646955. FAG acknowledges financial support from CONICYT through the project FONDECYT Regular Nr. 1181264, and from the Max Planck Society through a Partner Group grant.

%%%%%%%%%%%%%%%%%%%%%%%%%%%%%%%%%%%%%%%%%%%%%%%%%%


\bibliographystyle{mnras}
\bibliography{references} % if your bibtex file is called example.bib

%%%%%%%%%%%%%%%%%%%%%%%%%%%%%%%%%%%%%%%%%%%%%%%%%%

%%%%%%%%%%%%%%%%%%%%%%%%%%%%%%%%%%%%%%%%%%%%%%%%%%

% Don't these lines
\bsp	% typesetting comment
\label{lastpage}
\end{document}